\pdfoutput=1
\documentclass[aps,preprint]{revtex4}%
\usepackage{amsfonts}
\usepackage{amsmath}
\usepackage{amssymb}
\usepackage{graphicx}%
\providecommand{\U}[1]{\protect\rule{.1in}{.1in}}

\begin{document}
\preprint{ }

\begin{center}
{\LARGE Stochastic model for computer simulation}

{\LARGE of the number of cancer cells and lymphocytes }

{\LARGE in homogeneous sections of cancer tumors}

\bigskip

Arnulfo Castellanos Moreno, Alejandro Castellanos Jaramillo, Adalberto
Corella, Madue\~{n}o,

Sergio Guti\'{e}rrez L\'{o}pez, Rodrigo Arturo Rosas Burgos

Departamento de F\'{\i}sica, Universidad de Sonora, 83000, Hermosillo, Sonora,
MEXICO\textbf{\bigskip}

Abstract
\end{center}

We deal with a small enough tumor section to consider it homogeneous, such
that populations of lymphocytes and cancer cells are independent of spatial
coordinates. A stochastic model based in one step processes is developed to
take into account natural birth and death rates. Other rates are also
introduced to consider medical treatment: natural birth rate of lymphocytes
and cancer cells; induced death rate of cancer cells due to self-competition,
and other ones caused by the activated lymphocytes acting on cancer cells.
Additionally, a death rate of cancer cells due to induced apoptosis is
considered. Weakness due to the advance of sickness is considered by
introducing a lymphocytes death rate proportional to proliferation of cancer cells.

Simulation is developed considering different combinations of the parameters
and its values, so that several strategies are taken into account to study the
effect of anti-angiogenic drugs as well the self-competition between cancer
cells. Immune response, with the presence of a kind of specialized
lymphocytes, is introduced such that they appear once cancer cells are
detected. Induced apoptosis of cancer cells is introduced to model the action
of several drugs under development right now. Besides, the model predicts the
cancer relapse even from a very small number of cells. Simulation is done by
using Gillespie algorithm.\bigskip

Keyword: Noise; mathematical models; cancer; tumors.\bigskip

PACS: 02.50.Ey; 05.40.-a; 87.10.+e

\newpage

\section{Introduction}

The basic goal of this paper is to present a computer simulated model to study
the population time evolution of two interacting species: cancer cells and
lymphocytes, both confined in a very small tumor, such that spatial dependence
does not need to be considered. Several strategies to attack the disease are
simulated and two important conclusions are reached: the first one, the
efficiency is enhanced with a combination of anti-angiogenic drugs plus a
system based in artificial apoptosis of cancer cells; and the second one, a
small amount of cancer cells are enough for the reappearance of the disease.

Efforts to capture in a mathematical model the dynamic of chemical substances
and cells involved in the development of cancerous tumor started, at least,
since 1955 with a work about the cell structure of lung cancer under
radiotherapy \cite{Thom}. After 1980 the amount of papers dealing with cell
dynamics related to cancerous tumors rose significantly (See for example
\cite{Duchting1, Lefever, Adams, Wheldon}).

At first, a set of cancerous cells is an avascular tissue, without blood
vessels, and its dimension is just a few millimeters. Due to the lack of
nutrients, the internal shells of the tumor die, then, a segregation of
chemical substances produce vasculogenesis in the neighborhood of the small
tumor, such that new vessels appear connecting it with the cardiovascular
system. Once that food is available, the tumor grows to be an encapsulated
tumor or a malignant one with capacity to spread to neighboring organs. This
is known as metastasis.

Vascularization produces layers that randomly raise a barrier to block drugs
introduced inside the body to destroy cancerous cells. So, the application of
chemical substances becomes a spatial process where diffusion through a
chaotic tissue is very hard because the administered drug must survive for
internal mechanism used by living tissues disintoxicate themselves. For this
reason, it is difficult to penetrate a solid tumor preserving the level of
toxicity needed to kill cancerous cells \cite{Gavhane}.

Models to mathematically understand this problem has been directed to several
goals. One of these is to understand how a cytotoxic substance jumps the
barrier to reach cancerous cells \cite{Jain}. Another one is to study the
volumetric growth of the tumor after the action of chemotherapy and surgery
\cite{Rosell}. If someone is going to develop a mathematical model to study
tumors, a good advice is to take into account the classification presented by
T. Roose et. al. \cite{Roose}. They say that there are three different stages
for cancer development: avascular, vascular and metastatic. In the second one
the tumor has blood vessels, while in the first one has not. Once that there
is enough blood, solid tumors changes a lot of his properties and show several
pathological features. They are disorganized and its vasculature does not
work. In this cases mathematical models based on spatial diffusion processes
are needed (see for example Eugene J. Koay et. al.\cite{Koayetal}) Besides,
new approaches have been introduced to predict the size of tumors based on the
mitotic/apoptotic index and on diffusion penetration \cite{Edgertonetal}. In
the same sense, Jennifer Pascal et. al. have proposed a model based in a
differential equation of diffusion inside the tumor, obtaining a biophysical
description of the vascular/tissue architecture and drug perfusion within the
malignant zone \cite{Pascaletal}. In summary, metastatic stage appears when
cancer cells spread from the tumor to other parts of the body. Models for
avascular tumors are considered as a first step toward building models for
vascularized tumors. On the other hand, one should consider that when cancer
cells are in a high nutrient environment they proliferate, but in low nutrient
levels they die. Besides, there is an intermediate period called quiescent in
which nutrient levels do not permit proliferation but cancer cells stay alive.
Since any tumor needs blood to grow, a lot of research to find how to stop
angiogenesis (the process of building new blood vessels from the surrounding
body to the tumor) has been done. The nature of initial considerations can be
seen at \cite{Blagosklonny}. So, in 2008 Shaked et. al. \cite{Shaked}\ reached
the conclusion that clinical trials have indicated that drugs that inhibit the
growth of blood vessels can sometimes enhance the effectiveness of traditional
chemotherapy. For example, simultaneous administration of the antiangiogenic
drug bevacizumab with the chemotherapeutic agent paclitaxel improves survival
benefits for metastatic breast cancer and small cell lung cancer. This kind of
combined strategy suggests that it would be a good strategy to combine
chemotherapy with antiangiogenic drugs.

Based on these findings, we simplify our problem dealing with a tumor before
metastasis, so that population does not depend on spatial coordinates. As a
consequence, it is possible to work with a model based in computer simulations
with birth and death probabilities whose mathematical form takes into account
the number of cells acting in a small volume. This approach is similar to the
one consisting of working with particles in statistical physics to get
thermodynamics, fluctuations and out of equilibrium phenomena, including
systems with few particles like that of nanostructures. In a previous paper
\cite{Castellanos-et-al}\ we used the same hypothesis to develop an analytical
stochastic model where treatments based on chemotherapy were considered, and
the same mathematical approach can be generalized to deal with antiangiogenic
drugs, including a kind of artificially induced apoptosis in cancer cells. Two
coupled and nonlinear differential equations were found to describe the
temporal evolution of the average density of cancer cells and lymphocites.
Numerical solutions were obtained to explore the dynamical behavior of the
system. The existence of two basins was found in that paper: one where the
patient survives (health basin) plus other where he dies. Just as medical
evidence show us. However, by studying the random behavior around the
deterministic values, a possible noise-induced escape from health basin was
detected, because by evaluating the eigenvalues of the involved matrices, it
can be found that random fluctuations have unbounded standard deviations,
suggesting that disease could appear again. However, just linear noise was
considered because the analytical approach is based on van Kampen omega
expansion, and that kind of phenomena cannot be studied. A similar problem
appears if we want to deal with small populations. Therefore, we have
developed the present approach to explore that problem and to simulate the
combination of treatments based on different strategies.

This paper is organized as follows: a model of the interaction between
lymphocytes and cancer cells is presented in the second section to estimate
conditions needed to consider a stationary situation, such that the number of
both populations have the same number of individuals. Our attention is paid on
very small tumors (or a patch of a big one) to maintain independence of
spatial coordinates. A brief discussion about rate growth for cancer cells and
lymphocytes is presented in the third section to explain how the basic
parameters were selected. The model and the construction of transition rates
is presented in the fourth section, and also a summary of a first analytical
approach developed by the authors is included to show why computer simulations
are needed to go ahead in the understanding of the problem. Section five is
devoted to present the Gillespie algorithm, such that a reader with some
experience can reproduce this work. Computational results are presented in the
sixth section, and also the development with the absence and the presence of
specialized defense is considered and compared. On the other hand, the effect
of many no specialized lymphocytes is considered to explore what happen if the
process of maturity of lymphocytes does not give the tools to identify cancer
cells. Besides, strategies based on antiangiogenic drugs are studied too.
Results about released patients, declared free of disease under macroscopic
considerations, are shown to explain that it is true that random fluctuations
have unbounded standard deviations which can produce a patient suffers a
relapse. A combined strategy, such that antiangiogenic and artificially
induced apoptosis in cancer cells, is discussed to see that results are
better, preserving a better condition of patients. Finally, conclusions are presented.

\section{Materials and Methods}

In order to determine the size and the number of individuals to take into
account in the simulation, we consider the picture given by Macklin et. al.
\cite{Macklin}. They say that an avascular tumor can reach only a few
micrometers of diameter when it is feeding from the natural blood vessels of
the near region of the organism. By simulating the development of a tumor,
they found that a quiescent core forms at about 9 days, when the tumor radius
is approximately $0.34mm$, while a necrotic core appears around day 15 when
the radius of tumor is approximately $0.5mm$. Therefore, to deal with a lower
bound, in this paper we have considered a small tumor, such that its radius is
just $r_{T}\simeq200\mu m=0.2mm$. We assume that this tumor has a very small
central core of necrotized tissue due to the absence of blood irrigation. At
this stage of the tumor development, the available blood vessels are just the
usual density of capillaries in a healthy tissue, without any addition in
capacity to irrigate that part. When the amount of blood is not enough, the
cancer cells start to die due to the shortage of oxygen and nutrients. At the
same time the pressure executed by the growing tumor on the surrounding
healthy tissue destroys the extracellular matrix and changes the normal flux
of blood. Previous to metastasis, cancer cells produce chemical substances
such that the body starts the production of new vessels going to the tumor,
and once this goal is reached, a new stage is initiated where cancer cells
have a faster reproduction and their growing number is warrantied. Our goal is
to study the tumor before metastasis starts.

So, our model is developed by assuming several hypotheses: 1) there is a stage
where cancer cells are organized inside a small sphere, without the typical
chaotic growth of a tumor producing metastasis, such that cancer cells form a
battle line waiting for lymphocytes advancing against them through the normal
vessels of the body. 2) It is assumed that no shell exists yet, so that immune
system of the body can reach the cancer cells to destroy them. 3) Finally,
surviving cells and destroyed extracellular matrix are not considered between
cancer cells, such that they can be considered like little spheres inside the tumor.

Therefore, the estimation of the number of cells involved can be done by
taking into account geometric considerations. The volume of the blood inside
some part of the body can be estimated by using the next proportion: first, a
human body weighting $85kg$ has an average of $5%
\operatorname{l}%
$ of blood, so that we can assume that the percentage of blood in some place
of the tissue is of the order of $c=\frac{5}{85}=0.05882\,4$. On the other
hand, the blood flow speed in a body is such that at "aorta" has an average of
$40$ to $50\frac{cm}{s}$, while at capillaries is just as slow as
$v_{blood}=0.03\frac{%
\operatorname{cm}%
}{%
\operatorname{s}%
}=300\frac{%
\operatorname{\mu m}%
}{%
\operatorname{s}%
}$ (see for example \cite{Uzwiak}). With this assumption, we will get a
superior bound for time to be estimated, because a study done by K. Yamauchi
et. al. \cite{Yamauchi}\ have shown that cells migration through capillaries
is such that the minimum diameter to allow migration is $8%
\operatorname{\mu m}%
$, with a speed of cancer cells approximated to $48.3\frac{%
\operatorname{\mu m}%
}{%
\operatorname{h}%
}$. In the quicker cases, the biggest migration velocity was reached just by
20 cells with $13.2\frac{%
\operatorname{\mu m}%
}{%
\operatorname{h}%
}$.

Then, we consider a cylinder of area $A$ and height $h=v_{blood}\ast t$
containing a volume of blood given by%
\begin{equation}
V_{blood}=Av_{blood}tc \label{present1}%
\end{equation}

The number of leukocytes in the volume $V_{blood}$ can be found by considering
that its average density is%
\begin{equation}
5\ast10^{9}\frac{leukocytes}{%
\operatorname{l}%
}\leq\rho_{leu}\leq1.1\ast10^{10}\frac{leukocytes}{%
\operatorname{l}%
}. \label{present2}%
\end{equation}

Once these considerations are done, we can formulate the next question: how
many cancer cells are in touch with lymphocytes? To know that, we choose a
small spheroidal tumor whose radius is $r_{T}$, containing cancer cells with
radius $r_{c}$.%

\begin{center}
\includegraphics[
natheight=3.906400in,
natwidth=9.270800in,
height=2.0202in,
width=4.7565in
]%
{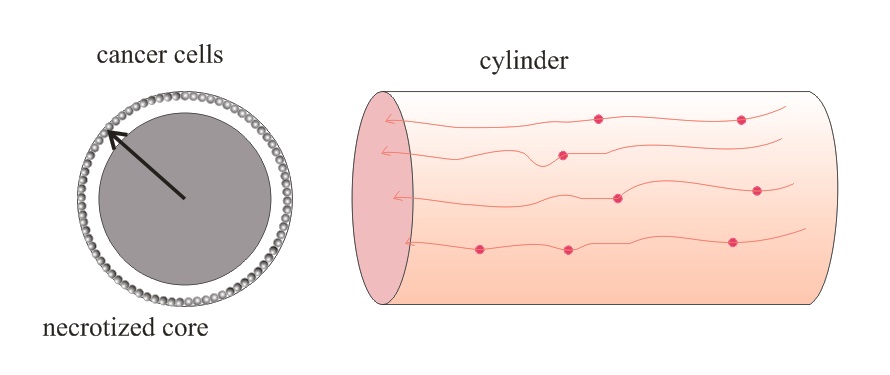}%
\\
Figure 1
\end{center}

The cross section of the tumor presented to blood flow is a half of the total
surface of the sphere, $A_{c}=4\pi r_{T}^{2}$, containing $N_{c}$ malignant
cells in its boundary in potential contact with blood. Since these cells
present a half of their surface to the blood flow, we obtain that%
\begin{equation}
N_{c}=\frac{\frac{1}{2}A_{T}}{\frac{1}{2}A_{c}}=\frac{\frac{1}{2}\left(  4\pi
r_{T}^{2}\right)  }{\frac{1}{2}\left(  4\pi r_{c}^{2}\right)  }=\left(
\frac{r_{T}}{r_{c}}\right)  ^{2}. \label{present3}%
\end{equation}
Besides, we need to answer the next question: how many lymphocytes can be in
touch with cancer cells? This quantity is estimated by assuming that blood
vessels going to the tumor are inside a cylinder whose area is $A=\pi
r_{T}^{2}$ and its volume is given by
\begin{equation}
V_{body}=Av_{b}t, \label{present4}%
\end{equation}
with $t$ a time which be determined ahead. The volume of blood contained in
this cylinder is%
\begin{equation}
V_{blood}=\left(  Av_{b}t\right)  c. \label{present5}%
\end{equation}
The amount of lymphocytes in the cylinder is%
\begin{equation}
N_{lin}=\left(  Av_{b}tc\right)  \rho_{leu}. \label{present6}%
\end{equation}
We consider a system whose initial conditions are $N_{lin}\left(  t=0\right)
=N_{c}\left(  t=0\right)  $, so that%
\begin{equation}
Av_{b}tc\rho_{leu}=\left(  \frac{r_{T}}{r_{c}}\right)  ^{2} \label{present7}%
\end{equation}
and%
\begin{equation}
t=\frac{\left(  \frac{r_{T}}{r_{c}}\right)  ^{2}}{Av_{b}c\rho_{leu}}.
\label{present8}%
\end{equation}
Since lymphocytes are present in the scenario as a component of blood flow,
and we have chosen to deal with an initial static situation at $t=0$, we need
to estimate the time $t$ such that both of the populations in competence are
numerically equal.

Such estimation of the time $t$ depends on the size of tumor and we have
selected the optimal scenario, the one where the state of evolution of the
tumor can be detected in a threshold where imaging based on magnetic resonance
is in its better resolution. In addition to that, we want to suggest a mechanism.

For example, we can propose periodic scanning of the chest of a woman, or the
prostate of a man, by using 1.5 Tesla MR scanner at a spatial resolution of
$117\times117\times300%
\operatorname{\mu m}%
$ \cite{Majumdar} and with micro-CT whose spatial resolution reach cubes with
$100\times100\times100%
\operatorname{\mu m}%
$ \cite{Badea}. It would be necessary a system to save and recover files with
old and new images, and techniques to compare them automatically. This can be
done with software based on neural networks.

Clearly, several approaches could be cited because work is being carried out
to detect cancer. For example, multiparametric magnetic resonance imaging is
considered as a promising method for the detection of prostate cancer
\cite{Sankinenietal}. Besides, a planar type antenna is reportedly a useful
design for detecting a cancerous tumor of 5 mm in size, in the early stages of
development \cite{Tiangetal}. On the other hand, Barbara Blasiak et. al.
\cite{Blasiaketal} have published a recent review where they discuss several
applications of nanomaterials for early and specific cancer detection and
therapy. And so on. The important point here is that very small tumors should
be considered.%

\[%
\begin{tabular}
[c]{cc}%
{\includegraphics[
natheight=6.312300in,
natwidth=5.219100in,
height=2.2373in,
width=1.8542in
]%
{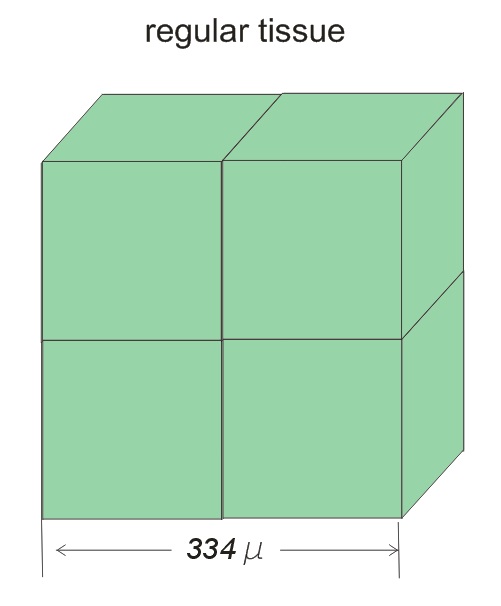}%
}
&
{\includegraphics[
natheight=6.072700in,
natwidth=7.937300in,
height=2.092in,
width=2.7268in
]%
{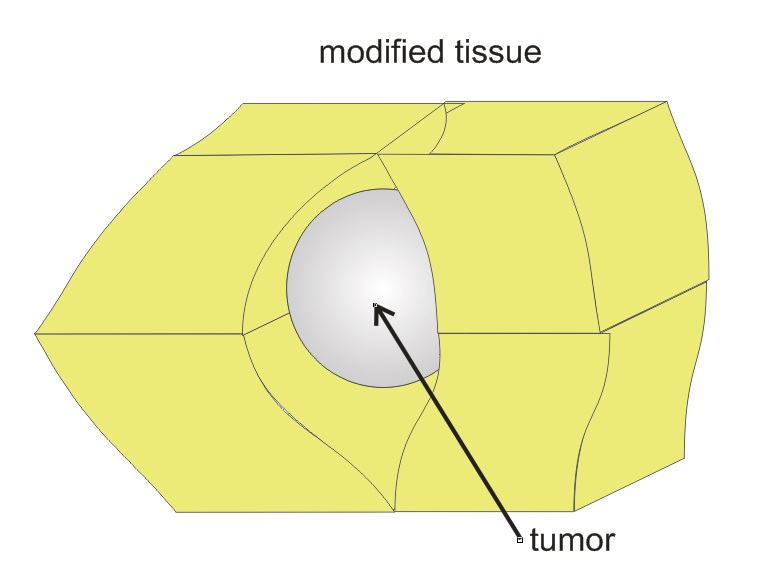}%
}
\\
\multicolumn{2}{c}{Figure 2}%
\end{tabular}
\
\]

To estimate $t$ we choose a tumor whose diameter is $200%
\operatorname{\mu m}%
$, and a typical cancer cell whose radius is $r_{c}=4%
\operatorname{\mu m}%
$. Then, the equation (\ref{present3}) gives us $N_{c}=625$. If $\rho
_{leu}=5\times10^{9}\frac{leukocytes}{\mu%
\operatorname{l}%
}$ we get%

\[
t=854.\,\allowbreak28%
\operatorname{s}%
,
\]
and with $\rho_{leu}=1.1\times10^{10}\frac{leukocytes}{\mu%
\operatorname{l}%
}$ we obtain%

\[
t=310.65%
\operatorname{s}%
,
\]
so that the interval to get the assumed scenario is%

\[
310.65%
\operatorname{s}%
<t<854.28%
\operatorname{s}%
.
\]
This is of the order of $5$ minutes to $14$ minutes and $15$ seconds. Since
the more aggressive cancer cells have a doubling time rate of two days, a
stationary situation where it is considered the same amount of cancer cells
and lymphocytes is an acceptable option.

In our model, extracellular matrix is considered just as an empty space to be
occupied, without taking into account that, due to its destruction by the
tumor, it is active in delivering nutrients to be used in the expansion of
malignant cells.

\subsection{Rate growth}

\subsubsection{Cancer cells reproduction speed}

The potential doubling time is defined as the time necessary to double the
number of proliferating tumor cells. According to the reference \cite{Hlatky},
the index named as potential doubling time ($T_{pot}$) is evaluated by using a
linear regression such that a straight line is fitted to write the number of
cells $N$ as follows:%
\[
N=A+Be^{K_{p}t},
\]
where $K_{p}$ is named the rate constant of cell production.

The number of counted cells $N\left(  t_{1}\right)  =N_{1}$ and $N\left(
t_{2}\right)  =N_{2}$, with $t_{2}>t_{1}$, are fixed as follows $N_{1}=10^{3}$
and $N_{2}=2\times10^{3}$, so that $T_{pot}=t_{2}-t_{1}$. Then, we have%
\[
T_{pot}=\frac{1}{K_{p}}\log\left(  \frac{2000-A}{B}\right)
\]
Taking $A=0$, we write down
\[
T_{pot}=\frac{1}{K_{p}}\log\left(  2\right)
\]

The next data can be found in (see Table 1 at \cite{Hlatky})%
\begin{align*}
&
\begin{tabular}
[c]{llll}%
tumor line & Origin & $T_{pot}$ days (*) & $T_{pot}$ days (**)\\
FaDu & SCC of pharynx & $2.0\pm0.2$ & $1.3\pm0.1$\\
HCT15 & Human colon adenocarcinoma & $2.5\pm0.1$ & $2.2\pm0.1$\\
STS26T & Schwannoma (soft tissue sarcoma) & $2.5\pm0.2$ & $2.4\pm0.2$\\
U87 & Glioblastoma multiforme & $3.9\pm0.3$ & $2.5\pm0.2$\\
SCC21 & SCC & $5.9\pm0.3$ & $2.9$\\
HGL9 & Glioblastoma multiforme & $6.2\pm0.3$ & $3.9\pm0.3$\\
U251-MG & Glioblastoma multiforme & $7.5\pm0.5$ & $10.5\pm1.4$%
\end{tabular}
\\
&  \text{Table 1}%
\end{align*}
where data marked with (*) are obtained with one of the methods described in
the last reference (named first method cytokinesis block), and data marked
with (**) are found by using another method explained in the same reference
(named there as second method IdUrd).

Then, we have evaluated the next data for the rate constant of cell
production, $K_{p}$, based in the expression: $K=\frac{\log\left(  2\right)
}{T_{pot}}$:%

\begin{align*}
&
\begin{array}
[c]{cc}%
T_{pot}\hspace{0.2cm}\left(  days\right)  & K_{p}\hspace{0.2cm}\left(
days^{-1}\right) \\
1 & 0.693147\\
1.5 & 0.462098\\
2 & 0.346574\\
2.5 & 0.277259\\
3 & 0.231049\\
3.5 & 0.198042\\
4 & 0.173287\\
4.5 & 0.154033\\
5 & 0.138629\\
5.5 & 0.126027\\
6 & 0.115525\\
6.5 & 0.106638\\
7 & 0.099021\\
7.5 & 0.0924196\\
8 & 0.0866434
\end{array}
\\
&  \text{Table 2}%
\end{align*}

Here, $K_{p}$ plays the same role as the parameter $b_{1}$ (to be introduced
ahead). So, we have assumed that an aggressive cancer tumor will be such that
\[
b_{1}=0.35
\]
where one cell produces a new one each two days.

\subsection{Lymphocytes reproduction speed}

According to reference \cite{Hellerstein}: "In healthy, HIV-1 seronegative
subjects, $CD4^{+}$ T cells had half-lives of 87 days and 77 days,
respectively, with absolute production rates of $10$ $CD4^{+}\frac{Tcells}{\mu%
\operatorname{l}%
}$\ per day and $6$ $CD8^{+}\frac{Tcells}{\mu%
\operatorname{l}%
}$\ per day."

Then, starting from $1$, a sequence obtained by adding the number $6$ during
$77$ times, reaches the value $463$. If we assume an exponential growth, it
must be a transition rate $b_{2}$ given by%
\[
b_{2}^{\prime}=\frac{1}{77}\log\left(  463\right)  =0.0797107,
\]
for $CD4^{+}$ lymphocytes, while a sequence such that the number $10$ is added
$88$ times give us%
\[
b_{2}^{\prime\prime}=\frac{1}{88}\log\left(  881\right)  =0.0770575,
\]
for $CD8+$ lymphocytes. Therefore, we have chosen the value%
\[
b_{2}=0.078
\]
for our simulations.

In normal conditions, a balance in the number of lymphocytes is found if it is
assumed that apoptosis is $d_{2}=0.078$, too.

\subsection{The model and a summary of a first analytical approach}

\subsubsection{How transition rates are built}

Our model for the tumor is a space where there are $\Omega$ small boxes, with
three categories for them: empty, occupied by a cancer cell, and occupied by a
lymphocyte; so that there are $M$ boxes occupied by lymphocytes, $N$ boxes
occupied by cancer cells, and $\Omega-M-N$ empty boxes. This is a collection
to be treated as a big recipient with $\Omega$ balls, with three possible
colors: a) a blue ball corresponds to an empty box, b) a white ball
corresponds to a lymphocyte, and c) a black ball corresponds to a cancer cell.
Events to be considered are classified in three groups: 1) to get just one
ball, 2) to find two balls, and 3) to obtain three balls from the big recipient.

We denote a white ball (lymphocyte) with the letter $B$, a black ball (cancer
cell) as $A$, and a blue ball (empty space) as $E$.

\begin{itemize}
\item In the first group of events we have: $B\rightarrow E$ (death of a
lymphocyte by apoptosis), and $A\rightarrow E$ (death of a cancer cell by
artificial apoptosis).

\item In the second group of events: $B+E\rightarrow B+B$ (birth of a
lymphocyte), $B+A\rightarrow B+B$ (death of a cancer cell due to the presence
of a lymphocyte, or specialized defense), $A+E\rightarrow A+A$ (birth of a
cancer cell), $A+A\rightarrow A+E$ (death of a cancer cell by self-competence).

\item In the third group we consider just one event: $A+A+B\rightarrow A+A+E$
(death of a lymphocyte due to weakness of the patient).
\end{itemize}

All of the events can occur with transition rates denoted as:

\begin{itemize}
\item Death rate (apoptosis rate) for lymphocytes: $d_{2}$, ($B\rightarrow E$).

\item Death rate (artificial apoptosis rate) for cancer cells: $d_{1}$,
$A\rightarrow E$.

\item Birth rate for lymphocytes: $b_{2}$, ($B+E\rightarrow B+B$).

\item Death rate for cancer cells; $c_{12}$, $B+A\rightarrow B+B$.

\item Birth rate for cancer cells: $b_{1}$, $A+E\rightarrow A+A$.

\item Death rate (self-competence) for cancer cells: $c_{11}$, $A+A\rightarrow
A+E$.

\item Death rate (weakness due to sickness) for lymphocytes: $c_{21}$,
$A+A+B\rightarrow A+A+E$.
\end{itemize}

Transition rates are built by using elementary probability, they are presented
in the right column of the Table 3.%
\begin{align*}
&
\begin{array}
[c]{cc}%
\text{event} & \text{transition rate}\\
B\rightarrow E & d_{2}\frac{M}{\Omega}\\
A\rightarrow E & d_{1}\frac{N}{\Omega}\\
B+E\rightarrow B+B & b_{2}\frac{M\left(  M-1\right)  }{\Omega\left(
\Omega-1\right)  }\\
B+A\rightarrow B+B & c_{12}\frac{MN}{\Omega\left(  \Omega-1\right)  }\\
A+E\rightarrow A+A & b_{1}\frac{N\left(  \Omega-N-M\right)  }{\Omega\left(
\Omega-1\right)  }\\
A+A\rightarrow A+E & c_{11}\frac{N\left(  N-1\right)  }{\Omega\left(
\Omega-1\right)  }\\
A+A+B\rightarrow A+A+E & c_{21}\frac{N\left(  N-1\right)  M}{\Omega\left(
\Omega-1\right)  \left(  \Omega-2\right)  }%
\end{array}
\\
&  \text{Table 3}%
\end{align*}

It will be seen ahead that the transition $B+A\rightarrow B+B$ is very
important because it involves death of a cancer cell due to the action of a
lymphocyte. In this case there are two actions in one, it is the substitution
of one $A$-cell by a $B$-cell. So, the number $N$ diminishes in $1$, while the
number $M$ increases in $1$. It is considered here as a specialized response
modeling the activation of the body defense system through a sequence of
biochemical events, such that this kind of white blood cell can identify
specific antigens through a molecule named TCR. It can recognize specific
antigens and can act as a cytotoxic cell against specific malignant cells in a
tumor in a very lethal way. This is the reason to name it as specialized defense.

\subsubsection{Analytical approach based on van Kampen expansion}

In \cite{Castellanos-et-al} we have developed an analytical model by defining
population densities as%
\begin{equation}
n=\frac{N}{\Omega}=\psi+\frac{1}{\sqrt{\Omega}}\eta,\hspace{0.2cm}m=\frac
{M}{\Omega}=\phi+\frac{1}{\sqrt{\Omega}}\xi, \label{analiticos 1}%
\end{equation}
with $\left(  \psi,\phi\right)  $ the deterministic part, and $\left(
\eta,\xi\right)  $ the random noise. Besides, we separated each transition
rate in a way similar to the one presented now%
\begin{equation}
D_{M}=d_{2}\frac{M}{\Omega}=d_{2}\phi+\frac{1}{\sqrt{\Omega}}d_{2}\xi=\Omega
D_{M}^{\left(  1\right)  }+\sqrt{\Omega}D_{M}^{\left(  0\right)  }.
\label{analiticos 2}%
\end{equation}
Thus, a van Kampen expansion was done and it was possible to separate the
problem in two of them. First, a deterministic one, where differential
equations must be solved to know $\left(  \psi\left(  t\right)  ,\phi\left(
t\right)  \right)  $. Second, a time dependent Ornstein-Uhlenbeck stochastic
process, whose density probability function obeys a Fokker-Planck equation
whose coefficients depend on the statistical properties of $\left\langle
\psi\right\rangle $ and $\left\langle \phi\right\rangle $.

The general approach presented there gives us a pair of coupled ordinary
differential equations for the deterministic part of the problem%
\begin{align}
\frac{d\left\langle \psi\right\rangle }{dt}  &  =f\left(  \left\langle
\psi\right\rangle ,\left\langle \phi\right\rangle ,\left\langle \psi^{j}%
\phi^{k}\right\rangle \right) \label{analiticos 3}\\
\frac{d\left\langle \phi\right\rangle }{dt}  &  =g\left(  \left\langle
\psi\right\rangle ,\left\langle \phi\right\rangle ,\left\langle \psi^{j}%
\phi^{k}\right\rangle \right) \nonumber
\end{align}
with $j$, $k=1,2$.

These equations can be studied numerically by assuming uncorrelation and very
small standard deviations in the deterministic part. These hypotheses can be
written as follows:%
\begin{equation}
\left\langle \psi\phi\right\rangle \simeq\left\langle \psi\right\rangle
\left\langle \phi\right\rangle ,\hspace{0.2cm}\left\langle \phi^{2}%
\right\rangle \simeq\left\langle \phi\right\rangle ^{2},\hspace{0.2cm}%
\left\langle \psi^{2}\right\rangle \simeq\left\langle \psi\right\rangle ^{2}
\label{analiticos 4}%
\end{equation}

Based in the analytical treatment, the next equations were found:%

\[
\frac{d\left\langle \psi\right\rangle }{dt}=b_{1}\left\langle \psi
\right\rangle -\left(  b_{1}+c_{12}\right)  \left\langle \psi\right\rangle
\left\langle \phi\right\rangle -\left(  b_{1}+c_{11}\right)  \left\langle
\psi^{2}\right\rangle -Q_{1}\left(  \left\langle \psi\right\rangle
+\varepsilon\left\langle \phi\right\rangle \right)
\]%
\[
\frac{d\left\langle \phi\right\rangle }{dt}=\left(  b_{2}-d_{2}\right)
\left\langle \phi\right\rangle -b_{2}\left\langle \psi\right\rangle
\left\langle \phi\right\rangle -b_{2}\left\langle \phi^{2}\right\rangle
-c_{21}\left\langle \psi^{2}\right\rangle \left\langle \phi\right\rangle
-Q_{1}\left(  \left\langle \psi\right\rangle +\varepsilon\left\langle
\phi\right\rangle \right)
\]

We have solved numerically this system of coupled nonlinear differential
equations and two basins were found: one where the patient survives (health
basin) plus other where he dies. We have accepted these results because the
well-known medical evidence is reproduced.

Besides, time evolution of the cancer cell density can be explored from the
first equation. To do that, we write it as%
\[
\frac{d\left\langle \psi\right\rangle }{dt}=b_{1}\left\langle \psi
\right\rangle \left\{  1-\frac{1}{K_{L}}\left\langle \phi\right\rangle
-\frac{1}{K_{c}}\left\langle \psi\right\rangle \right\}  -Q_{1}\left(
\left\langle \psi\right\rangle +\varepsilon\left\langle \phi\right\rangle
\right)
\]
where%
\[
K_{c}=\frac{1}{1+\frac{c_{11}}{b_{1}}}
\]
and%
\[
K_{L}=\frac{1}{1+\frac{c_{12}}{b_{1}}}
\]

Logistic growth is the particular case where there is no chemotherapeutic
treatment and lymphocytes are not present. For this reason we put $Q_{1}=0$,
and $\left\langle \phi\right\rangle =0$. Then
\[
\frac{d\left\langle \psi\right\rangle }{dt}=b_{1}\left\langle \psi
\right\rangle \left\{  1-\frac{1}{K_{c}}\left\langle \psi\right\rangle
\right\}
\]
and a stationary state can be reached because $\lim_{t\rightarrow\infty
}\left\langle \psi\right\rangle _{s}=K_{c}$. Therefore $K_{c}$ is the carrying
capacity of the population of cancer cells.

However, lymphocytes are an important part of the system and they play a
fundamental role in the defense of the patient. So that it is necessary to see
what happens in the healthy basin. This can be explored by assuming a case
where successful chemotherapeutic treatment has been applied and retired with
good results. We put $Q_{1}=0$, and $\left\langle \phi\right\rangle _{s}%
\simeq1$, because almost all the available spaces are occupied by lymphocytes.
Now the solutions are%
\[
\left\langle \psi\right\rangle _{s}^{\left(  1\right)  }\simeq0
\]
and%
\[
\left\langle \psi\right\rangle _{s}^{\left(  2\right)  }\simeq K_{c}\left(
1-\frac{1}{K_{L}}\right)  =-\frac{c_{12}}{b_{1}}%
\]
But densities cannot be negative, so that $\left\langle \psi\right\rangle
_{s}^{\left(  2\right)  }$ is not acceptable. As a consequence, from a
macroscopic point of view, like in medical diagnosis, health has been recovered.

However there is a hidden problem. In the stochastic case densities are%
\[
\psi=\left\langle \psi\right\rangle +\eta,\hspace{0.3cm}\phi=\left\langle
\phi\right\rangle +\xi
\]

Denoting the random fluctuations as $\left(  \eta,\xi\right)  =\left(
q_{1},q_{2}\right)  =\vec{q}$, the time dependent Ornstein-Uhlenbeck process
obeys the following Fokker-Planck equation%
\begin{equation}
\frac{\partial\Pi\left(  \vec{q},t\right)  }{\partial t}=-%
{\displaystyle\sum\limits_{\mu=1}^{2}}
\frac{\partial\left[  A_{\mu}\left(  \left\langle \psi\right\rangle
,\left\langle \phi\right\rangle ,\vec{q}\right)  \Pi\left(  \vec{q},t\right)
\right]  }{\partial q_{\mu}}+\frac{1}{2}%
{\displaystyle\sum\limits_{\mu=1}^{2}}
{\displaystyle\sum\limits_{\nu=1}^{2}}
\frac{\partial^{2}\left[  D_{\mu\nu}\left(  \left\langle \psi\right\rangle
,\left\langle \phi\right\rangle ,\vec{q}\right)  \Pi\left(  \vec{q},t\right)
\right]  }{\partial q_{\mu}\partial q_{\nu}} \label{analiticos 5}%
\end{equation}
with
\begin{equation}
\left\{  A_{\mu}\right\}  =\left(
\begin{array}
[c]{c}%
A_{1}\\
A_{2}%
\end{array}
\right)  =\mathbf{L}\left(  \left\langle \psi\right\rangle ,\left\langle
\phi\right\rangle \right)  \vec{q} \label{analiticos 6}%
\end{equation}
the flux term and $\mathbf{L}\left(  \left\langle \psi\right\rangle
,\left\langle \phi\right\rangle \right)  $\ the matrix convection. The
diffusion matrix is given as%
\begin{equation}
\left\{  D_{\mu\nu}\right\}  =\left(
\begin{array}
[c]{cc}%
D_{11}\left(  \left\langle \psi\right\rangle ,\left\langle \phi\right\rangle
\right)  & D_{12}\left(  \left\langle \psi\right\rangle ,\left\langle
\phi\right\rangle \right) \\
D_{12}\left(  \left\langle \psi\right\rangle ,\left\langle \phi\right\rangle
\right)  & D_{22}\left(  \left\langle \psi\right\rangle ,\left\langle
\phi\right\rangle \right)
\end{array}
\right)  =\mathbf{D} \label{analiticos 7}%
\end{equation}
Since the random part depends on the deterministic one, the process is named a
nonautonomous system. Correlations are defined as%
\begin{equation}
\left\{  \Xi_{\mu\nu}\right\}  =\left\{  \left\langle q_{\mu}q_{\nu
}\right\rangle -\left\langle q_{\mu}\right\rangle \left\langle q_{\nu
}\right\rangle \right\}  =\mathbf{\Xi}, \label{analiticos 8}%
\end{equation}
and satisfy the differential equations written by van Kampen \cite{van-Kampen}%
:%
\begin{equation}
\frac{d\mathbf{\Xi}}{dt}-\mathbf{L\Xi-\Xi L}=\mathbf{D} \label{analiticos 9}%
\end{equation}

The solution to (\ref{analiticos 5}) is a Gaussian function given as%
\begin{equation}
\Pi\left(  \vec{q},t\right)  =\frac{1}{\sqrt{2\pi}\sqrt{\det\mathbf{\Xi}}}%
\exp\left\{  -\frac{1}{2}\left[  \vec{q}-\left\langle \vec{q}\right\rangle
\left(  t\right)  \right]  \cdot\mathbf{\Xi}^{-1}\left[  \vec{q}-\left\langle
\vec{q}\right\rangle \left(  t\right)  \right]  \right\}
\label{analiticos 10}%
\end{equation}

Using that $\Xi_{12}=\Xi_{21}$, the vector $\vec{z}$ is defined as%
\[
\vec{z}=\left(
\begin{array}
[c]{c}%
\Xi_{11}\\
\Xi_{12}\\
\Xi_{22}%
\end{array}
\right)
\]
and the system (17) can be written as
\[
\frac{d\vec{z}}{dt}=\mathbf{A}\vec{z}+\vec{d}%
\]
with $\mathbf{A}$\ a matrix whose components depend on time through the pair
$\left\{  \left\langle \psi\right\rangle ,\left\langle \phi\right\rangle
\right\}  $. The trivector $\vec{d}$\ contains the diffusion. This expression
has been used to analyze the case $\left\langle \phi\right\rangle _{s}\simeq1$
and $\left\langle \psi\right\rangle _{s}\simeq0$, founding that the
eigenvalues of the matrix $\mathbf{A}$ take positive values; then, the
time-evolution of $z_{i}\left(  t\right)  $ is not bounded. This suggest that
a possible noise-induced escape from health basin could be hidden here.

In summary, the analytical approach give us the next conclusions:

\begin{enumerate}
\item The deterministic part is divided in two basins: one of healthy patient,
with final states near $\left(  0,1\right)  $; and other one for death
patients, such that the final state is near $\left(  1,0\right)  $.

\item The random part is such that fluctuations could maintain instability in
final states, such that healthy patients are in risk.
\end{enumerate}

However, this analytical approach cannot help us to explore the last
conclusion, because these time dependent Ornstein-Uhlenbeck stochastic
processes are useful for small fluctuations, and since these goes like $\sim
N^{-\frac{1}{2}}$, small populations result unreachable. So, we have to
consider that this simulation approach is necessary to include any number of
cancer cells and lymphocytes and without the hypothesis (\ref{analiticos 4}).

\subsection{Algorithm for computer simulation}

Once transition rates and events are available, Doob and Gillespie algorithm
can be applied to get the simulations \cite{Gillespie-libro}. This was done
taking into account a software developed by Th. Newman \cite{Newman-paper,
Newman}. The original problem in Gillespie paper consists in a chemical system
confined in a volume $V$ and evolving on time, such that it can be specified
by $M$ state variables $h_{1}$, $h_{2}$, ..., $h_{M}$, with $h_{\mu}$ being
the number of different molecular reactants combinations for each reaction
$R_{\mu}$. ($\mu=1,...,M$). Obviously, there should be transition rates
specified (please see \cite{Gillespie-paper}). In our paper, these transition
rates are built by using elementary probability (see Table 3).

Gillespie approach considers a reaction probability density function $P\left(
\tau,\mu\right)  $, such that $P\left(  \tau,\mu\right)  d\tau$ is the
probability, at time $\tau$, that a reaction will occur in the interval
$\left(  t+\tau,t+\tau+d\tau\right)  $. He followed a reasoning similar to the
one expressed to study the probability of collisions in gases \cite{Reif}.
First, he considered the probability $P_{0}\left(  \tau\right)  $ that no
reaction occurs in the interval $\left(  t,t+\tau\right)  $. Second, it was
taken into account the probability\ $h_{\mu}c_{\mu}d\tau$, to first order in
$d\tau$, that a reaction $R_{\mu}$ occurs in the interval $\left(
t+\tau,t+\tau+d\tau\right)  $. If the no overlapping intervals are
statistically independents, the probability that there is no reaction in the
interval $\left(  t,t+\tau\right)  $, but there is a reaction at $\left(
t+\tau,t+\tau+d\tau\right)  $, is
\begin{equation}
P\left(  \tau,\mu\right)  d\tau=P_{0}\left(  \tau\right)  h_{\mu}c_{\mu}%
d\tau\label{sim 1}%
\end{equation}
Then, he found that the density probability $P\left(  \tau,\mu\right)  $ is
given as%
\begin{equation}
P\left(  \tau,\mu\right)  =h_{\mu}c_{\mu}e^{-\sum_{\nu=1}^{M}h_{\nu}c_{\nu
}\tau} \label{sim 2}%
\end{equation}

As a consequence, a stochastic process described by the transitions given at
Table 3 can be simulated, according to Gillespie, with the following
algorithm. Here it is named as a one urn model of no spatial cancer cells and
lymphocytes:\bigskip

Step = $0$: Initialization

The number $\Omega$ of objects in the urn is generated. This is the number of
sites in the line of battle of our paper.

The number of realizations $N_{0}~$is specified. $N_{0}=1$ in the historical
mode, or $N_{0}\sim10^{4}$ in the final mode.

Transition rates are introduced: $\sigma_{i}$, $i=1,..,M$, with $M$ the number
of interactions between two species.

Initial time $t=0$\ is set.

The initial number of cancer cells $N_{A}\left(  0\right)  $, and lymphocytes
$N_{B}\left(  0\right)  $, are specified. Store $\left(  t,N_{A},N_{B}\right)
$.

The variable $sigma\_sum=0$ is set.\bigskip

Step = $1$: Iteration

\begin{enumerate}
\item Produces an actualization of the variable $sigma\_sum$ as follows:%
\begin{equation}
sigma\_sum=%
{\displaystyle\sum\limits_{i=1}^{M}}
\sigma\left(  i\right)  \label{sim 3}%
\end{equation}

\item Generates two random numbers $r_{1}$ and $r_{2}$, with homogeneous
probability in $\left[  0,1\right]  $.

\item Creates time increment%
\begin{equation}
\tau=\frac{1}{sigma\_sum}\ln\left(  \frac{1}{r_{1}}\right)  \label{sim 4}%
\end{equation}
This gives us a random $\tau$ with exponential distribution.

\item Actualizes the time $t\rightarrow t+\tau$.

\item Random number $r_{2}$\ is used to choose one of the $M$ interactions.

\item Actualizes $N_{A}$ and $N_{B}$ according to the event chosen in 5.

\item Stores $\left(  t+\tau,N_{A},N_{B}\right)  $ and go to 1.
\end{enumerate}

\section{Results and Discussion}

Our simulation of the line of battle in an avascular tumor, described above,
was developed considering the following fixed parameters: $b_{1}=0.35$,
$b_{2}=d_{2}=0.078$, and $c_{12}$, $c_{21}$, $c_{11}$; and $d_{1}$ changing
such that real behavior reproduced in medical practice can be simulated.
However, since we are dealing with the simulation in a line of battle, an
estimation of real time cannot be obtained at this stage of the model. We must
recognize that this is a problem under consideration right now, for example,
some authors have presented reasons to think that development of big tumors is
exponential at first, but polynomial after that \cite{Fadia}.

The simulation produces the next results:

\subsection{Checking the algorithm}

At first, the approach was checked by working with $b_{1}=d_{1}=0.35$, and the
others parameters equal to zero. Initial conditions where: $\Omega=6250$,
$N\left(  t=0\right)  =625$ and $M\left(  t=0\right)  =1$. Algorithm was run
in two modes: in the first one the time evolution is saved to a file
(historical mode), and in the second one just the final data are saved to a
file once a time has been reached (final results mode). The historical mode is
presented in figure $3$.%

\begin{center}
\includegraphics[
natheight=8.979400in,
natwidth=10.520400in,
height=2.8115in,
width=3.2889in
]%
{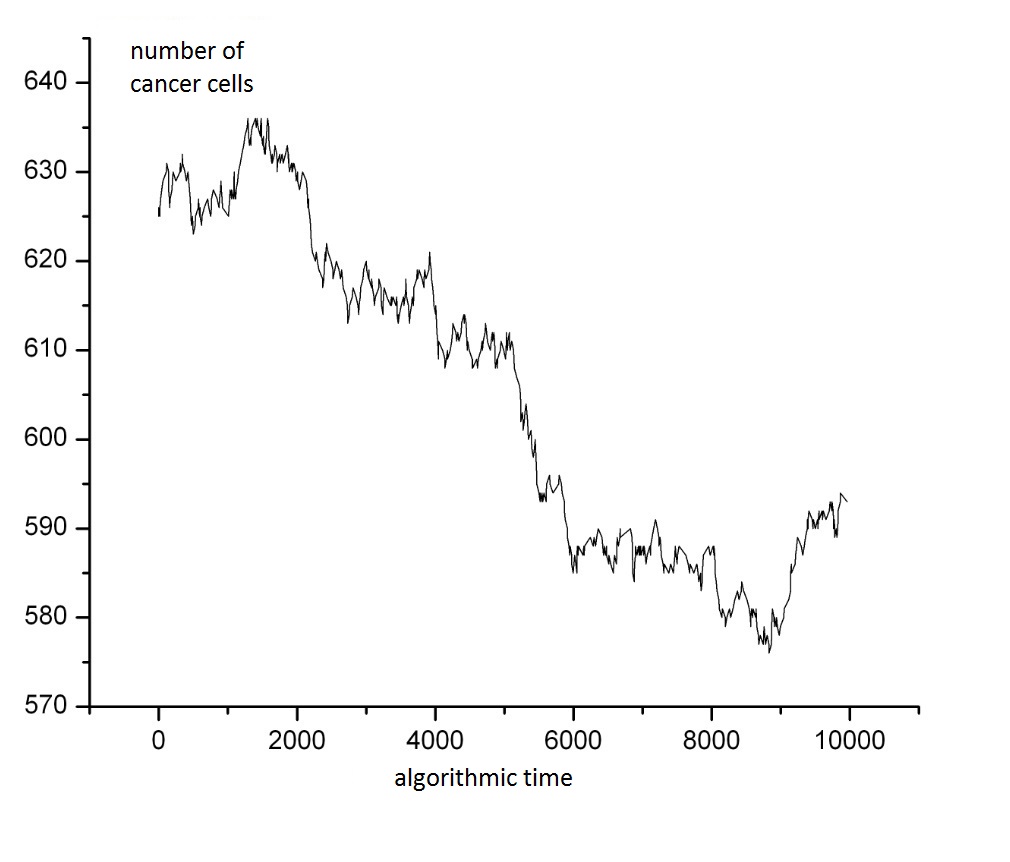}%
\\
Figure 3
\end{center}

In the final results mode $10^{4}$ data were generated, then, mean and
standard deviation were evaluated to obtain:%
\[
\left\langle N\right\rangle =591.625,\hspace{0.2cm}\hspace{0.2cm}%
\hspace{0.2cm}\sigma_{N}=23.9466,
\]
where $\left\langle N\right\rangle $ and $\sigma_{N}$\ are taken on the sample
of data obtained.

A similar work was done to see the behavior of lymphocytes when $b_{2}%
=d_{2}=0.078$, and the rest of parameters equal to zero. Here the initial
conditions were: $\Omega=6250$, $N\left(  t=0\right)  =1$, and $M\left(
t=0\right)  =625$. The historical mode gives us a graph presented in figure
$4.$%

\begin{center}
\includegraphics[
natheight=9.333100in,
natwidth=10.583600in,
height=2.9213in,
width=3.3088in
]%
{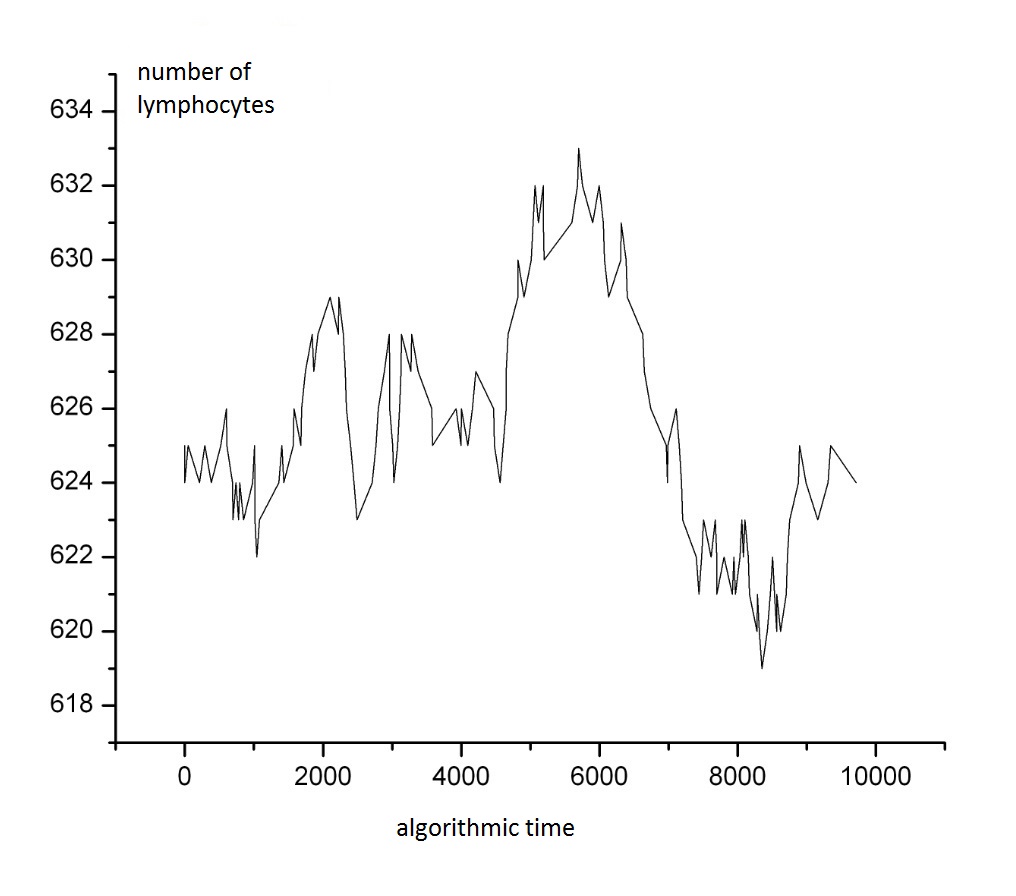}%
\\
Figure 4
\end{center}

After that, the final results mode was used to generate $10^{4}$ data. Mean
and standard deviation gave us:%
\[
\left\langle M\right\rangle =617.166,\hspace{0.2cm}\hspace{0.2cm}%
\hspace{0.2cm}\sigma_{M}=12.0828
\]

\subsection{Development without specialized defense}

Here we will consider two different systems. In one of them, sentinel
macrophages are activated when tissue homeostasis is perturbed, so that
soluble mediators are released, like cytokines, chemokines, matrix remodeling
proteases and reactive oxygen species, bioactive mediators such as histamine
to induce mobilization and infiltration of additional leukocytes into damaged
tissue. The other one is a more sophisticated adaptive immune system developed
through direct interaction with mature antigen-presenting cells where the
adaptive immune cells are involved, like B lymphocytes, $CD4^{+}$ helper T
lymphocytes and $CD8^{+}$ cytotoxic lymphocytes \cite{Karin-et-al}. Both
systems, specialized, and non-specialized defense, are simulated ahead.

First we simulate the case of a sick patient without medical treatment, this
is done by taking the following initial conditions: $\Omega=6250$, $N\left(
t=0\right)  =M\left(  t=0\right)  =625$. Regime of a sick person\ without
medical attention was simulated by using the parameters: $b_{1}=0.35$,
$b_{2}=d_{2}=0.078$, with $c_{12}=c_{21}=c_{11}=d_{1}=0$. Historical mode
gives us the graph shown in figure $5$.%

\begin{center}
\includegraphics[
natheight=9.791400in,
natwidth=10.458200in,
height=3.0623in,
width=3.2707in
]%
{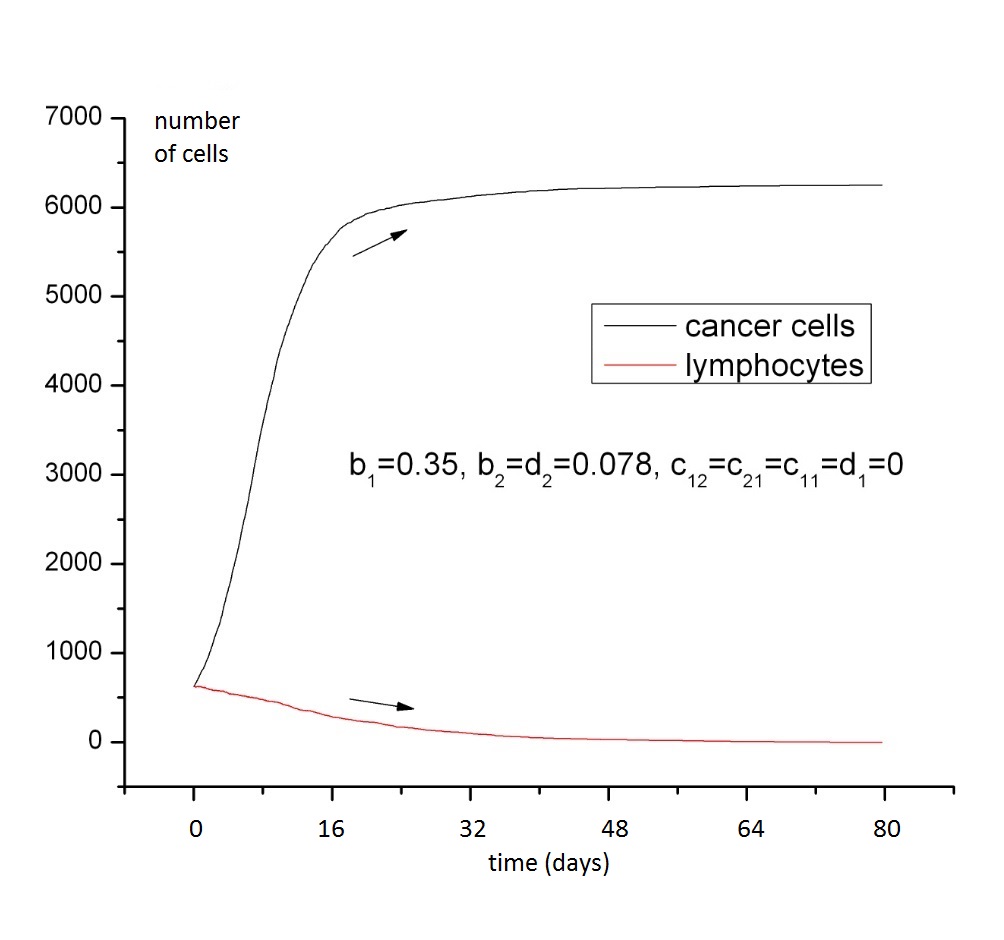}%
\\
Figure 5
\end{center}

Using the final data mode, we developed $10^{4}$ realizations to evaluate the
mean and the standard deviation of cancer cells in the time needed to take the
number of lymphocytes down to zero. This is%
\[%
\begin{array}
[c]{ccc}
& \text{Number} & \text{Volume }\left(
\operatorname{\mu m}%
\right)  ^{3}\\
\left\langle N\right\rangle  & 6248.72 & 1.67517\times10^{6}\\
\sigma_{N} & 0.596 & 1.598
\end{array}
\]
Average time and its standard deviation is presented in algorithmic time and
its translation to days is done by dividing by the number of available spaces
$\Omega=6250$.%
\[%
\begin{array}
[c]{ccc}
& \text{Algorithmic time} & \text{days}\\
\left\langle t\right\rangle  & 537\hspace{0.05cm}657 & 86.02\\
\sigma_{t} & 63\hspace{0.05cm}530 & 10.16
\end{array}
\]
\bigskip

This can be interpreted in the sense that cancer cells prevail because the
cancer cells of the line of battle are successful to produce, in $86$ days, a
growing tumor whose extension is added in a volume of $1.67517\times
10^{6}\left(
\operatorname{\mu m}%
\right)  ^{3}$. Obviously the action of lymphocytes is not enough.

\subsection{Development with immune response (specialized defense)}

The immune specialized response is simulated by considering the same initial
conditions used in the previous subsection. It is a regime of a sick
person\ without medical attention but with good immune response, and the
parameters used were: $b_{1}=0.35$, $b_{2}=d_{2}=0.078$, with $c_{12}%
=1\times10^{-5}$, and $c_{21}=c_{11}=d_{1}=0$.

Historical mode gives us the graph shown in figure 6:%

\begin{center}
\includegraphics[
natheight=9.000100in,
natwidth=10.895800in,
height=2.7268in,
width=3.2958in
]%
{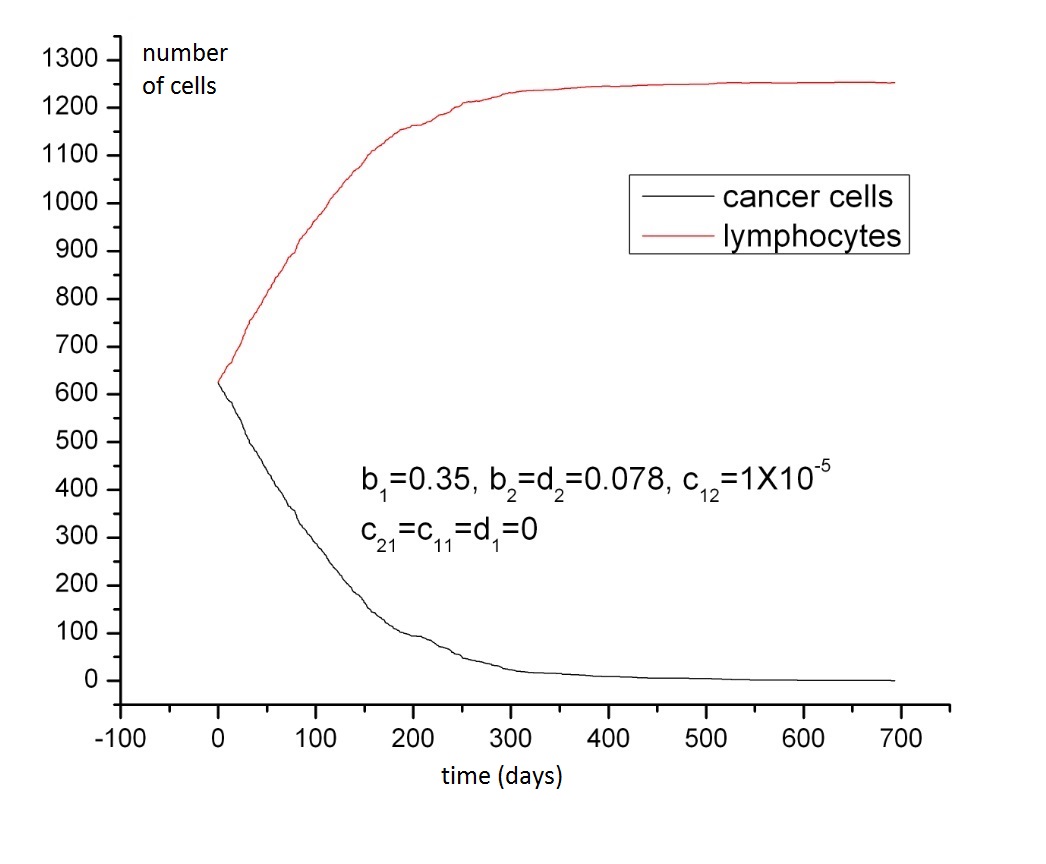}%
\\
Figure 6
\end{center}

We find that by increasing the specialized defense against cancer cells, the
lymphocytes prevail on the line of battle in just $300$ unites of algorithmic
time. If one has $25$ layers of cancer cells, $7500$ units of algorithmic time
would be enough. This result suggests three reflections:

\begin{enumerate}
\item According to this model, colonies of cancer cells could appear on the
organism frequently, but they would be destroyed before a tumor can reach a
critical radius to start metastasis.

\item A tumor grows enough, and appears with a critical radius, due to the
absence of a specialized defense where the enemies (cancer cells) are clearly
identified and a specific kind of lymphocytes are produced and directed against.

\item A tumor with a critical radius appears just when a barrier is opposed to
blood flux and lymphocytes do not get touch with cancer cells.
\end{enumerate}

The final data mode was developed by simulating $10^{4}$ realizations to
evaluate the mean and the standard deviations of lymphocytes in the time
needed to take the number of cancer cells down to zero. It was found that%
\[%
\begin{array}
[c]{cc}
& \text{Number}\\
\left\langle M\right\rangle  & 1251.34\\
\sigma_{M} & 4.35
\end{array}
\]%
\[%
\begin{array}
[c]{ccc}
& \text{Algorithmic time for 25 layers} & \text{days}\\
\left\langle t\right\rangle  & 14\hspace{0.05cm}840.9 & 2.374\\
\sigma_{t} & 2\hspace{0.05cm}474.47 & 0.396
\end{array}
\]

In attention to the third reflection, we can cite a review by Tr\'{e}dan et.
al. \cite{Tredan}. They have discussed that the effectiveness of drug therapy
is impaired by limited delivery of drugs to some regions of tumors. So, we can
think that similar results should be true for the action of specialized
defense of the body, such that if cancer cells survive, it is because they are
not in touch with lymphocytes.

\subsection{Development with specialized defense and a weak patient}

We consider the same condition in immune specialized response, but now with a
patient whose production of lymphocytes is diminished due to proliferation of
cancer cells. This was simulated by using the parameters: $b_{1}=0.35$,
$b_{2}=d_{2}=0.078$, $c_{12}=1\times10^{-5}$, and $c_{21}=1$, while
$c_{11}=d_{1}=0$.

Historical mode gives us the graph of figure $7$.%

\begin{center}
\includegraphics[
natheight=8.833200in,
natwidth=10.978800in,
height=2.6783in,
width=3.3217in
]%
{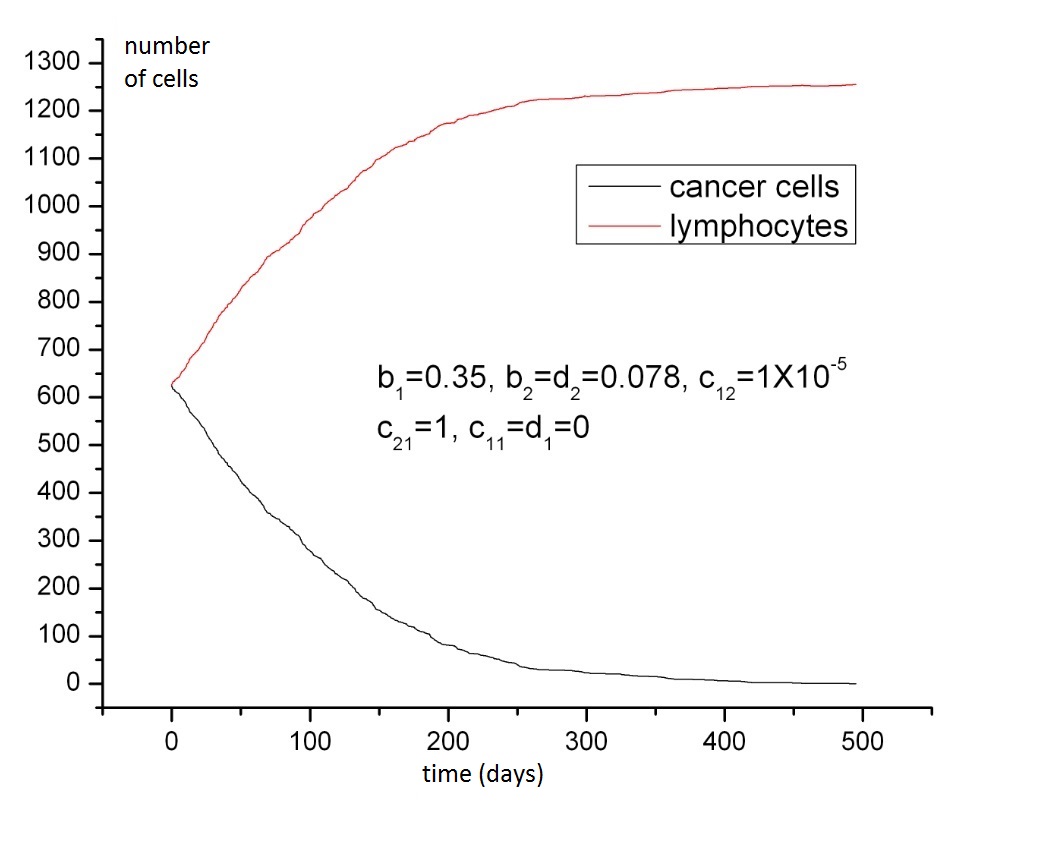}%
\\
Figure 7
\end{center}

It is found that specialized defense against cancer cells gives good results,
even in the case of weakness in patients. This is corroborated by using the
final data mode. We simulate $10^{4}$ realizations to obtain the mean and the
standard deviations of lymphocytes in the time needed to take the number of
cancer cells down to zero. The results are%
\[%
\begin{array}
[c]{cc}
& \text{Number}\\
\left\langle M\right\rangle  & 1251.34\\
\sigma_{M} & 4.35
\end{array}
\]%
\[%
\begin{array}
[c]{ccc}
& \text{Algorithmic time for 25 layers} & \text{days}\\
\left\langle t\right\rangle  & 14\hspace{0.05cm}858 & 2.378\\
\sigma_{t} & 2\hspace{0.05cm}507.93 & 0.401
\end{array}
\]

\subsection{Development with a lot of lymphocytes but non-specialized defense}

This is a strategy where reproduction rate of lymphocytes grow but in a
general sense, without acting through the transition: $B+A\rightarrow B+B$.
This is done with the following values of the parameters: $b_{1}=0.35$,
$b_{2}=1$, $d_{2}=0.078,$ $c_{12}=c_{21}=c_{11}=d_{1}=0$. Results are shown in
figure $8$.%

\begin{center}
\includegraphics[
natheight=8.770900in,
natwidth=10.937300in,
height=2.6584in,
width=3.3088in
]%
{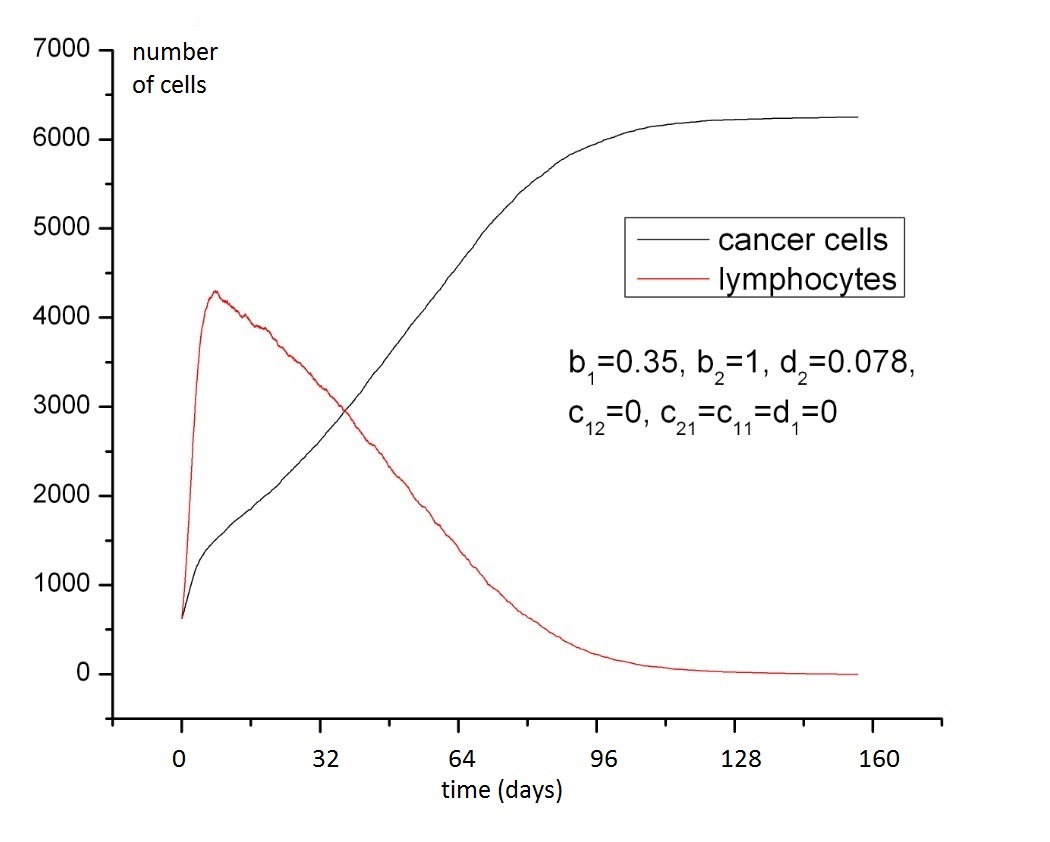}%
\\
Figure 8
\end{center}

With the parameters $b_{1}=0.35$, $b_{2}=1.5$, $d_{2}=0.078,$ $c_{12}%
=c_{21}=c_{11}=d_{1}=0$, the graph obtained is shown in figure 9.%

\begin{center}
\includegraphics[
natheight=9.083100in,
natwidth=11.208000in,
height=2.6628in,
width=3.2785in
]%
{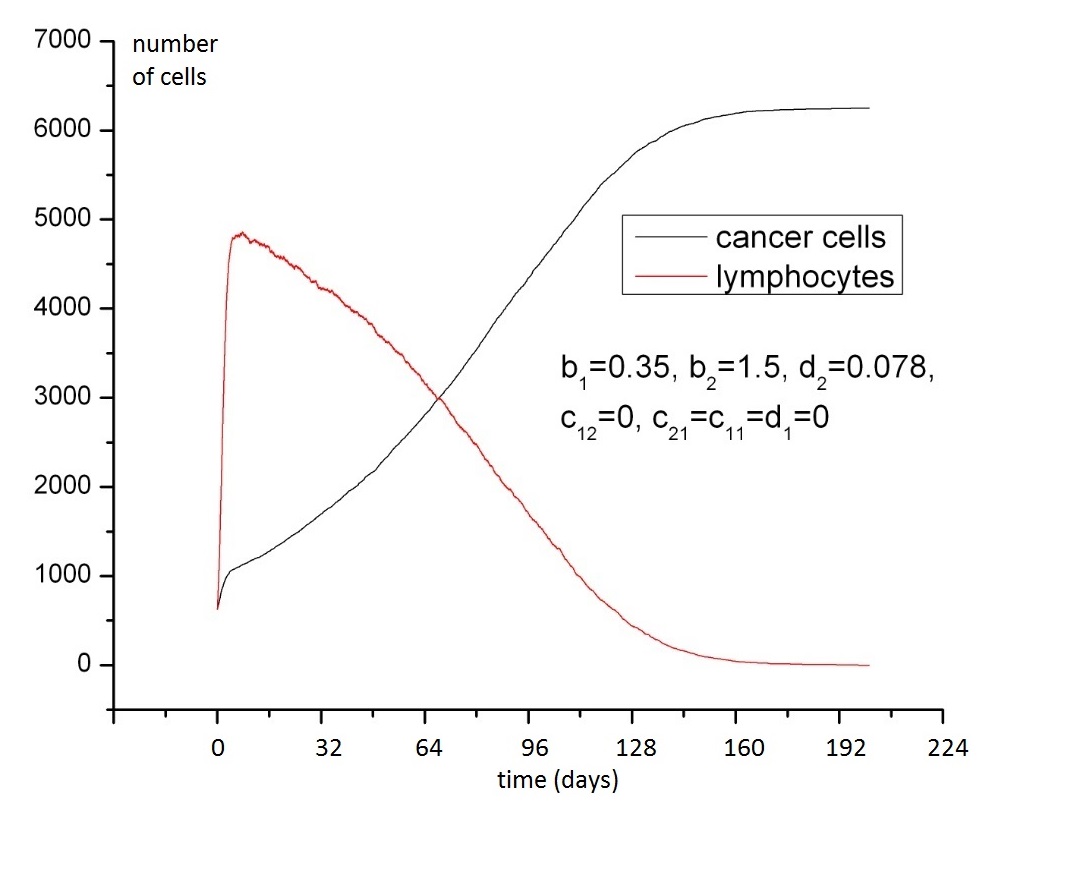}%
\\
Figure 9
\end{center}
And with the parameters $b_{1}=0.35$, $b_{2}=2$, $d_{2}=0.078,$ $c_{12}%
=c_{21}=c_{11}=d_{1}=0$, the results are shown in figure $10$.%

\begin{center}
\includegraphics[
natheight=8.760500in,
natwidth=10.812700in,
height=2.655in,
width=3.2707in
]%
{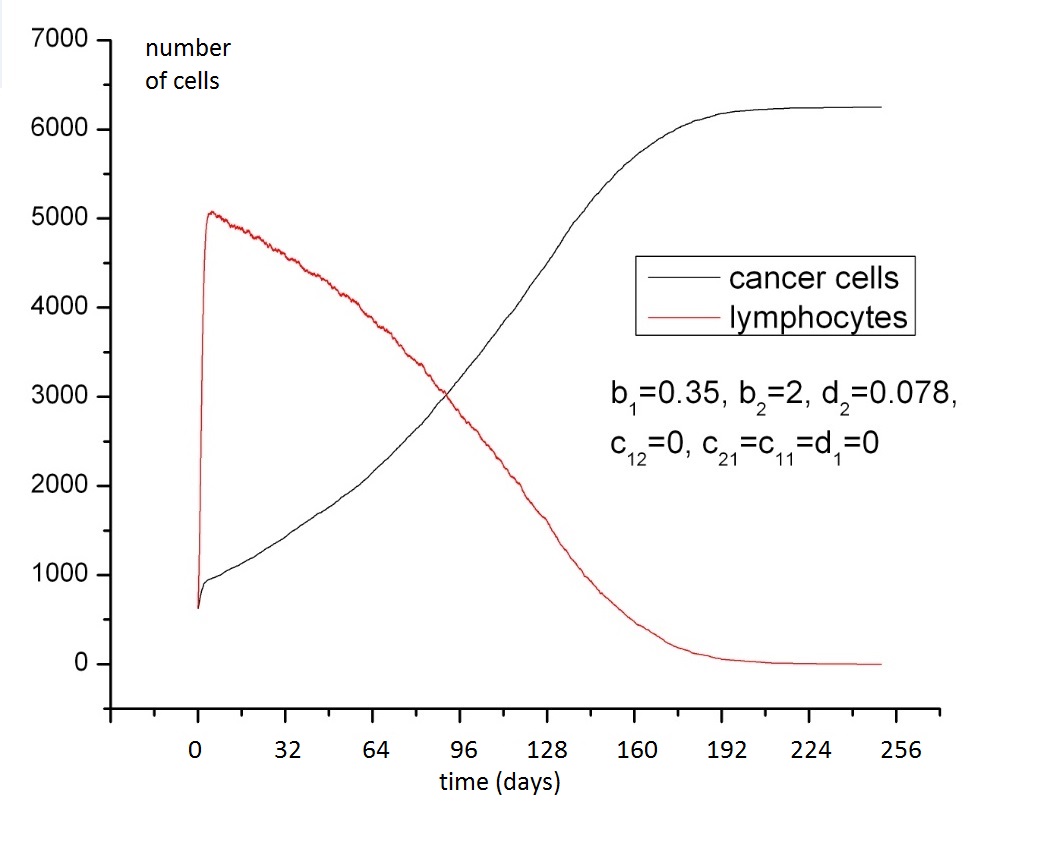}%
\\
Figure 10
\end{center}
We found that even with a big amount of lymphocytes, good results are not
reached. The absence of specialized lymphocytes produces fatal results.

\subsection{Strategy based on anti-angiogenic drugs. self-competition in
cancer cells}

A new broad family of anticancer drugs has been approved in recent years.
These are angiogenesis inhibitors administrated to avoid the building of new
blood vessels going from the main blood flow to the region where the tumor has
born. They can inhibit the production of angiogenic proteins, neutralize
angiogenic proteins or inhibit receptors where these proteins are received.
Some drugs induce endothelial cell apoptosis.

The general effect of these new drugs is to restrict, in some way, the flow of
oxygen and nutrients to the tumor. So, this is analog to the situation of a
colony of bacteria in a recipient with a finite capacity of oxygen and
nutrients. This is modeled in an approach named "logistic growth", where a
quadratic term is added ad hoc to introduce a control in the, otherwise,
exponential growth. The number of bacteria in the recipient is stabilized
because they are competing for food.

This is the case in cancer cells when they are in the core of a tumor, so that
blood flow does not arrive there and necrotized tissue appears as a
consequence. However, in the line of battle is totally different, because
there is a lot of blood there, so that reproduction is possible. Our idea is
to avoid all the complex details involved in the process produced by
anti-angiogenic drugs, to model their effect as one where oxygen and nutrients
are not enough, even in the surface of the tumor. So, the effect of
antiangiogenic drugs is modeled through the constant $c_{11}$, as
self-competition between cancer cells looking for oxygen and nutrients running
out. It is the transition: $A+A\rightarrow A+E$.

Working with the parameters $b_{1}=0.35$, $b_{2}=d_{2}=0.078$, $c_{12}%
=c_{21}=d_{1}=0$, and $c_{11}=1\times10^{-5}$, and using the historical mode,
we found the graph shown in figure $11$.%

\begin{center}
\includegraphics[
natheight=9.375400in,
natwidth=10.708100in,
height=2.8392in,
width=3.2404in
]%
{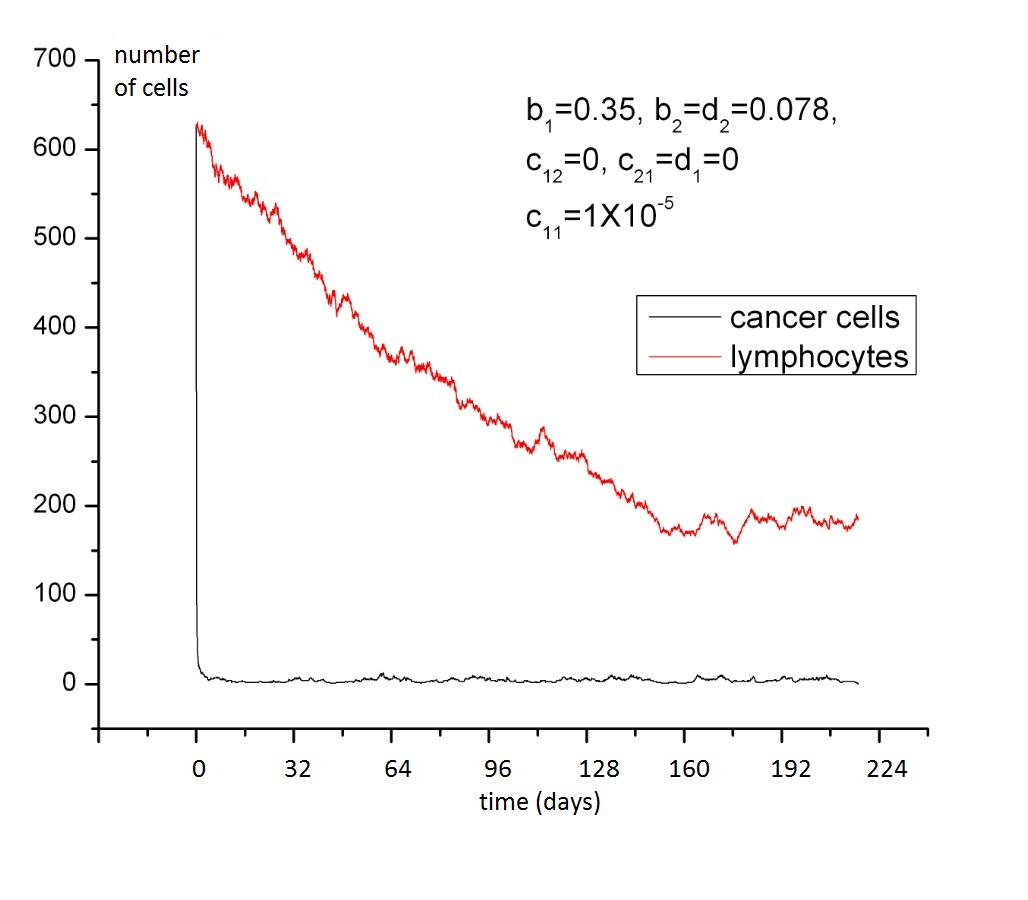}%
\\
Figure 11
\end{center}

Figure 11 is very important because contains two interesting results:

First: we can see that number of cancer cells in the line of battle goes to
zero, however, the amount of lymphocytes goes down to a very dangerous level.
This is an unexpected consequence because any transition rate of this model is
considering side effects. We have found in the literature that anti-angiogenic
therapy could cause several toxicities and biological effects, please see
reference \cite{ChungWu}.

Second: cancer cells go to zero after a very long time with very small random
fluctuations. At a first glance, medical treatment can be retired because
macroscopic manifestations cannot be detected. However, this is precisely the
regime that we could not solve with our analytical approach. A regime with
unbounded standard deviations. We will see in the next subsection what happens
in this case.

\subsection{Released patients under macroscopic considerations}

By observing figure 11 we could think that it is better to withdraw a
treatment when the number of lymphocytes is of the order of $400$, before they
reach a dangerous level. We have modeled that option by starting from the
initial conditions $N\left(  t=0\right)  =3$ and $M\left(  t=0\right)  =625$,
with the parameters $b_{1}=0.35$, $b_{2}=d_{2}=0.078$, with $c_{12}%
=c_{21}=c_{11}=d_{1}=0$. The historical mode gives us the results shown in
figure $12$.%

\begin{center}
\includegraphics[
natheight=8.937800in,
natwidth=10.708100in,
height=2.7095in,
width=3.2404in
]%
{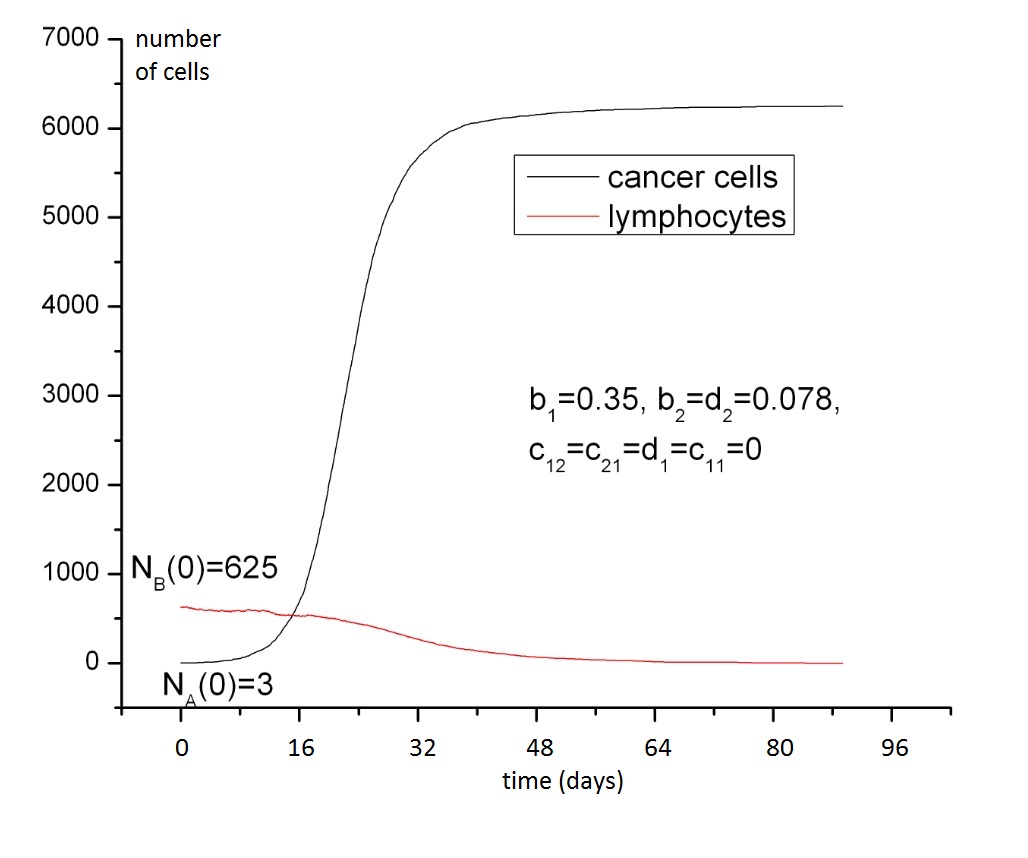}%
\\
Figure 12
\end{center}

Even for small values of the number of cancer cells, we found fatal results.
So, an important prediction of the model is that there is only one option: to
take the number of cancer cells down to zero.

In this dormant state there is no clinical or radiological evidence, because
the number of malignant cells is very low, thus its reproduction and organic
dissemination rates are low too. But there already are mutations associated
with malignant processes that are detected with ultrasensitive molecular
methods. See for example \cite{Newmanetal}.

\subsection{Induced apoptosis in cancer cells}

Induced apoptosis of cancer cells is introduced with the constant $d_{1}$ in
order to model the action of the Cuban drug Vidatox 30 CH, as reported by A.
D\'{\i}az et. al. \cite{Diazetal}. They wrote that venom of the scorpion
Rhopalurus Junceus caused the cellular death through the apoptosis in tumor
cells. It is the transition $A\rightarrow E$, and was simulated with the
parameters $b_{1}=0.35$, $b_{2}=d_{2}=0.078$, with $c_{12}=c_{21}=c_{11}=0$,
and $d_{1}=\left\{  0.1,0.2,0.3,0.4,0.5\right\}  $. The historical mode gives
us the results shown in figures 13 to 17.%

\begin{center}
\includegraphics[
natheight=8.291800in,
natwidth=10.104500in,
height=2.6809in,
width=3.2603in
]%
{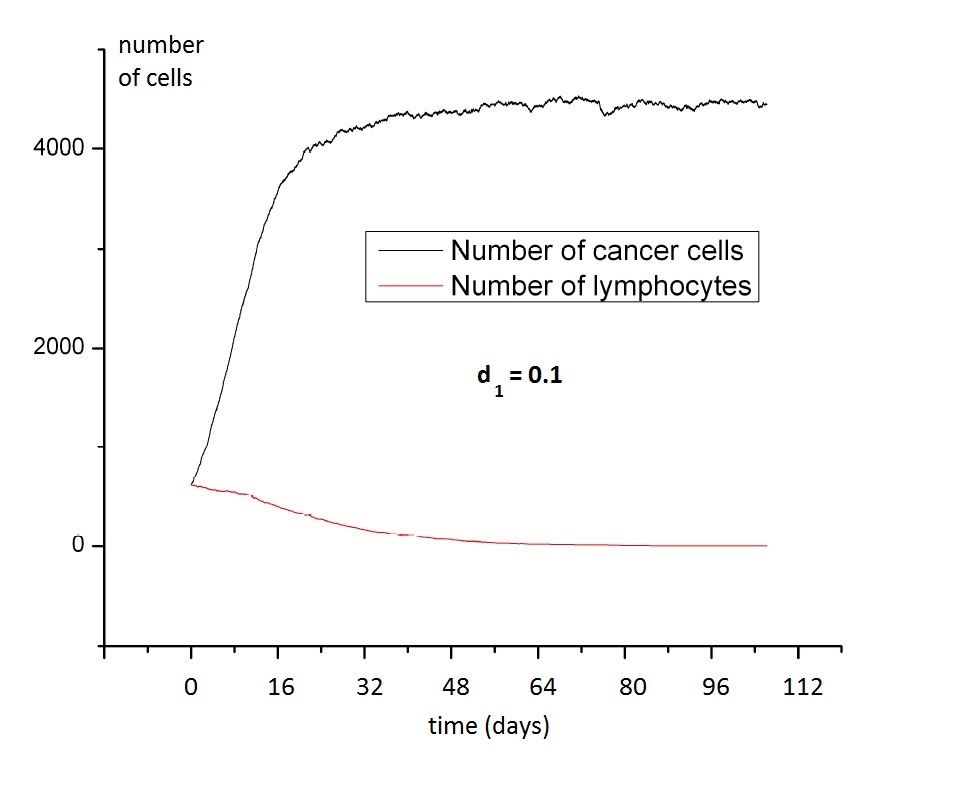}%
\\
Figure 13
\end{center}
%

\begin{center}
\includegraphics[
natheight=8.062700in,
natwidth=9.854500in,
height=2.6878in,
width=3.2794in
]%
{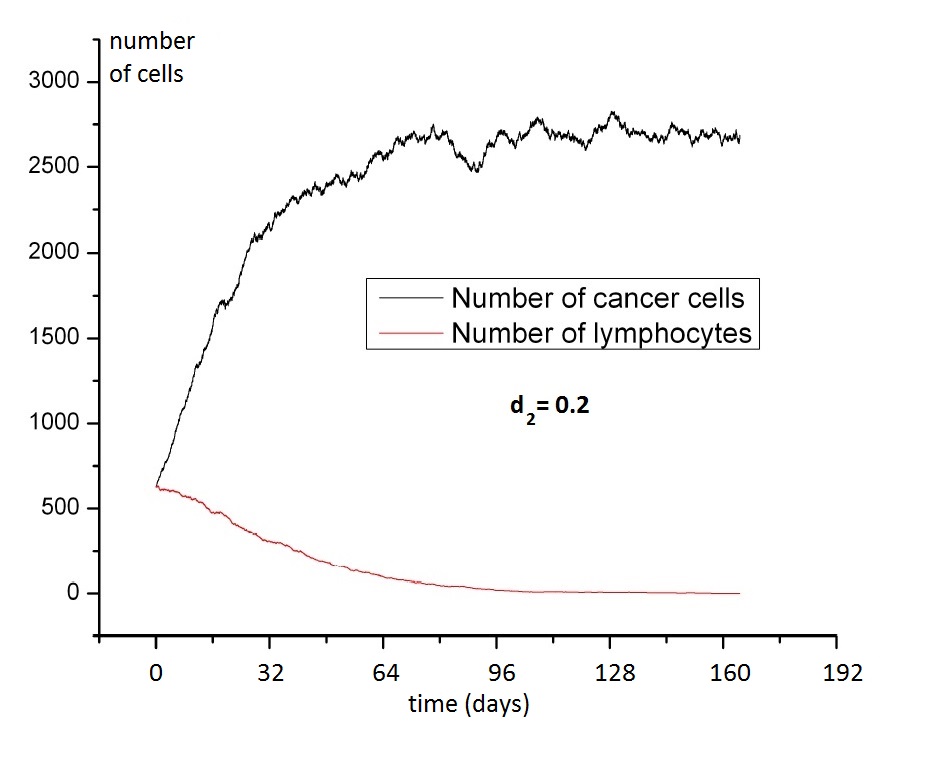}%
\\
Figure 14
\end{center}
%

\begin{center}
\includegraphics[
natheight=7.895700in,
natwidth=9.146300in,
height=2.8703in,
width=3.32in
]%
{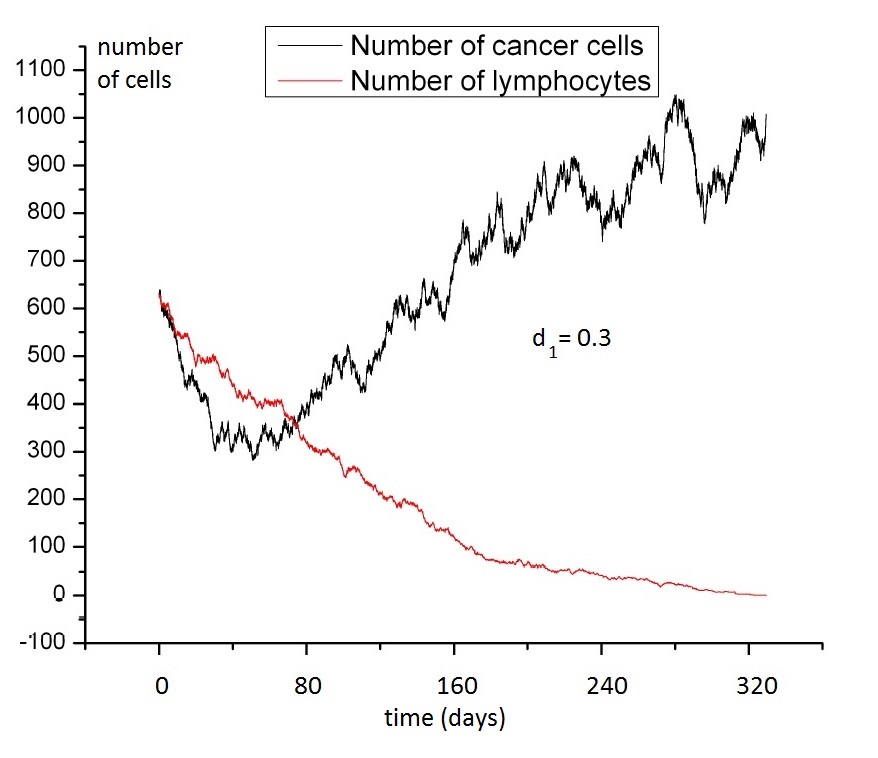}%
\\
Figure 15
\end{center}
%

\begin{center}
\includegraphics[
natheight=8.395600in,
natwidth=9.479200in,
height=2.8816in,
width=3.25in
]%
{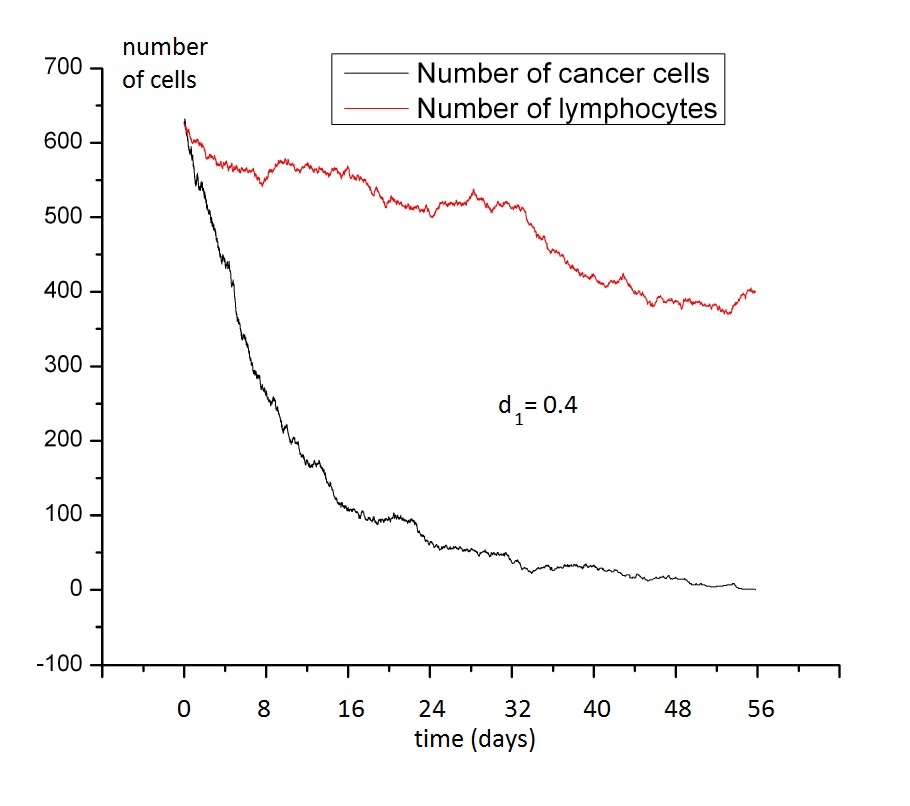}%
\\
Figure 16
\end{center}
%

\begin{center}
\includegraphics[
natheight=8.271100in,
natwidth=9.353800in,
height=2.9231in,
width=3.3018in
]%
{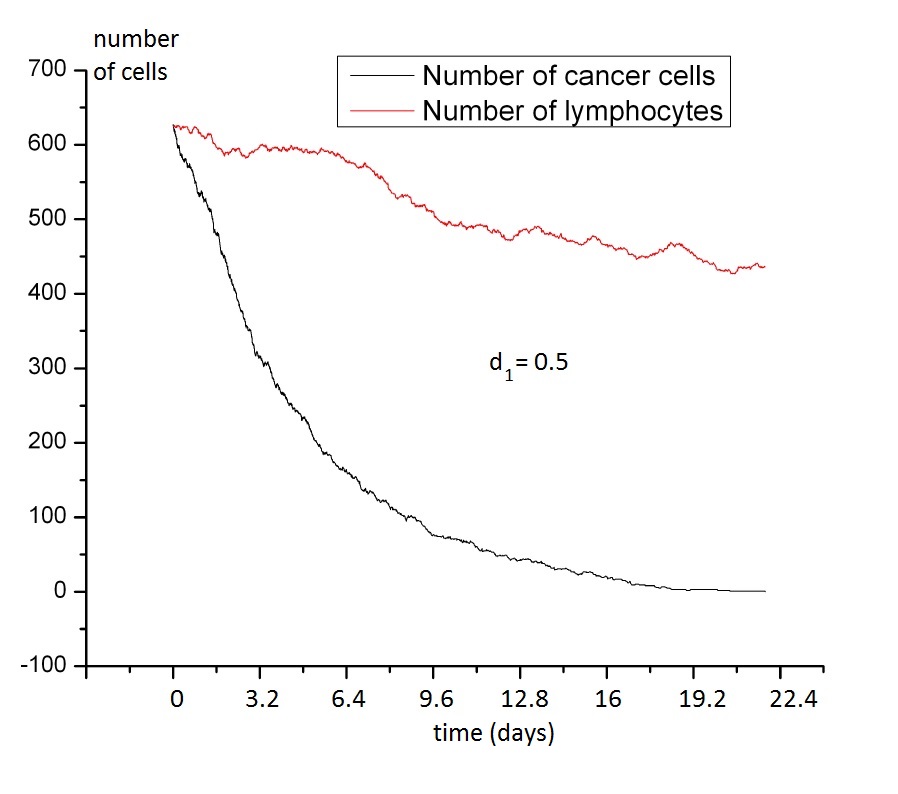}%
\\
Figure 17
\end{center}

We have found that induced apoptosis gives us good results provided that the
parameter takes higher values. It can be seen that the amount of lymphocytes
is higher than $400$ with the values of the parameter $d_{1}=0.4,0.5$.

\subsection{Combining anti-angiogenic drugs and induced apoptosis in cancer
cells}

Now we can review what happen if anti-angiogenic and inducer apoptosis drugs
are combined.

The historical mode gives us the results shown in figure 18.%

\begin{center}
\includegraphics[
natheight=8.145700in,
natwidth=9.416900in,
height=2.8781in,
width=3.3235in
]%
{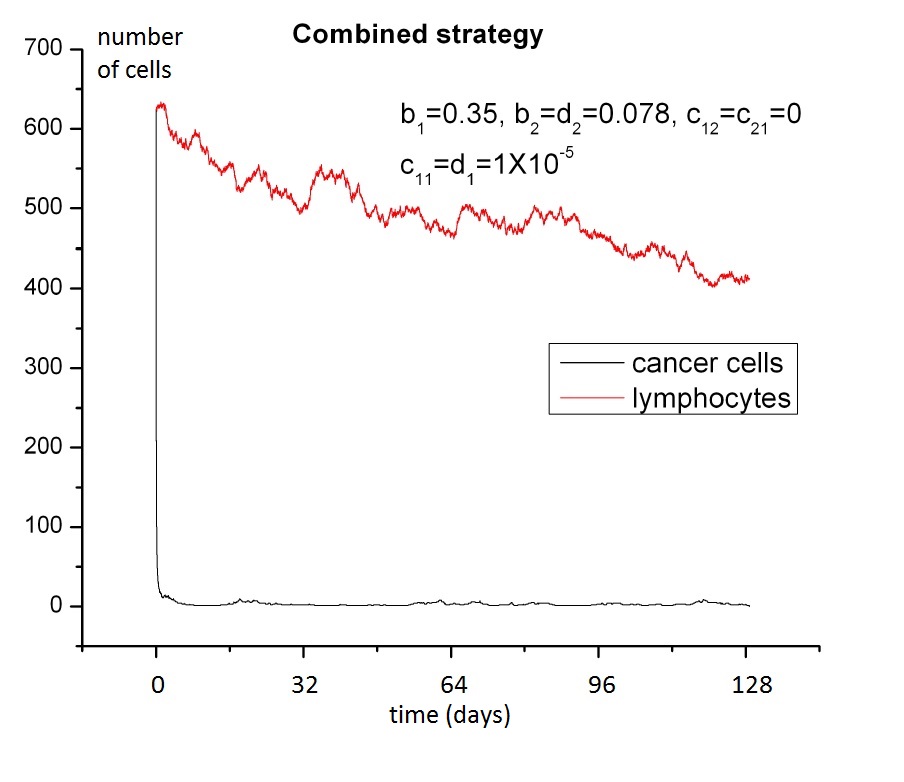}%
\\
Figure 18
\end{center}

Here the number of cancer cells goes to zero, when the number of lymphocytes
is higher than the obtained in the strategy based on anti-angiogenic drugs. A
specific detail was analyzed by using the final data mode: we developed
$10^{4}$ realizations with the parameters $b_{1}=0.35$, $b_{2}=d_{2}=0.078$,
with $c_{12}=c_{21}=0$, and $c_{11}=1\times10^{-5}$, $d_{1}=0.1$. As was
presented above, the results with $c_{11}=1\times10^{-5}$ are positive, but
with a dangerous amount of lymphocytes. Besides, $d_{1}=0.1$ does not produce
good results. On the contrary, if both of them are combined, the number of
cancer cells goes to zero with the following mean and standard deviations for
time and number of lymphocytes:
\[%
\begin{array}
[c]{cc}
& \text{Number}\\
\left\langle M\right\rangle  & 501.18\\
\sigma_{M} & 80.26
\end{array}
\]%
\[%
\begin{array}
[c]{cc}
& \text{Algorithmic time for 1 layer}\\
\left\langle t\right\rangle  & 218\hspace{0.05cm}320\\
\sigma_{t} & 170\hspace{0.05cm}917
\end{array}
\]
It can be seen that standard deviation is too large, $78.29\%$ of the mean. As
we have found before, this kind of random behavior is higher in the change
from fatal results to healthy patients. To review that, we developed $10^{4}$
realizations in the final data mode, with the parameters: $b_{1}=0.35 $,
$b_{2}=d_{2}=0.078$, with $c_{12}=c_{21}=0$, but $c_{11}=1\times10^{-4}$,
$d_{1}=0.2$. After the statistical analysis we found%
\[%
\begin{array}
[c]{cc}
& \text{Number}\\
\left\langle M\right\rangle  & 613.65\\
\sigma_{M} & 15.86
\end{array}
\]%
\[%
\begin{array}
[c]{cc}
& \text{Algorithmic time for 1 layer}\\
\left\langle t\right\rangle  & 14\hspace{0.05cm}663.1\\
\sigma_{t} & 8\hspace{0.05cm}941.89
\end{array}
\]
The number of lymphocytes is higher in $22.44\%$ than in previous case;
besides, time is shorter and its standard deviation has diminished to
$60.98\%$ of the mean time.

Then, one important conclusion of this work is that efficiency is enhanced
with a combination of anti-angiogenic drugs plus a system based in artificial
apoptosis of cancer cells.

While some reticence against induced-apoptosis in cancer cells exists, as an
option to obtain successful anti-cancer drugs, it must be recognized that it
is a mechanism strongly considered right now. Favorable opinions can be found
in \cite{Majidetal}.

\section{Conclusions}

The results obtained in this work allow us to discuss the following:

\begin{itemize}
\item This model predicts that cancer can relapse from a very small number of
cancer cells due to the existence of unbounded standard deviations.

\item Specialized immune response of a living being is definitely crucial in
fighting cancerous tumors.

\item Even in the case of a weakened patient, sufficient production of
specialized leukocytes may be decisive in the survival of the organism.

\item When the body does not have specialized leukocytes, its capacity to deal
with the problem of a cancerous tumor is very limited, and the prognosis is
unfortunately death.

\item The use of anti-angiogenic drugs is a strategy that can be used to
remove cancerous tumors. However, they also affect the rest of the body; they
can weaken and place in a fragile state to the patient before the attack of
other possible diseases. Furthermore, if the cancer cells are not eliminated,
there is a high probability that they re-grow in number, and again the result
is fatal.

\item A strategy based on a novel drug that causes apoptosis of cancer cells
shows great potential in fighting cancerous tumors, but it also produces some
weakening of the body.

\item The combination of strategies seems to be a better alternative because
it could produce a very rapid decrease in cancer cells, reducing the amount of
medication to be administered and therefore would be less side effects.
\end{itemize}

\end{document}